\documentclass[]{emulateapj}

\usepackage{enumerate}

\begin{document}

\newcommand{\suzaku}{{\it Suzaku} }
\newcommand{\integral}{{\it INTEGRAL} }
\newcommand{\rxte}{{\it RXTE} }
\newcommand{\asca}{{\it ASCA} }
\newcommand{\rosat}{{\it ROSAT} }
\newcommand{\einstein}{{\it Einstein} }
\newcommand{\chandra}{{\it Chandra} }
\newcommand{\xmm}{{\it XMM-Newton} }
\newcommand{\swift}{{\it Swift} }
\newcommand{\ginga}{{\it Ginga} }
\newcommand{\sax}{{\it BeppoSAX} }
\newcommand{\red}{\textcolor{red}}
\newcommand{\green}{\textcolor{green}}
\newcommand{\blue}{\textcolor{blue}}
\newcommand{\erg}{\mathrm{erg}}
\newcommand{\eV}{\mathrm{eV}}
\newcommand{\mwd}{M_{\mathrm{WD}}}
\newcommand{\rwd}{R_{\mathrm{WD}}}
\newcommand{\zfe}{Z_{\mathrm{Fe}}}
\newcommand{\msun}{M_\odot}
\newcommand{\zsun}{Z_\odot}
\newcommand{\nel}{n_{\mathrm{e}}}
\newcommand{\mel}{m_{\mathrm{e}}}
\newcommand{\persec}{\mathrm{s}^{-1}}
\newcommand{\arcminsq}{\mathrm{arcmin}^{-2}}
\newcommand{\cmsq}{\mathrm{cm}^{-2}}
\newcommand{\cmcubed}{\mathrm{cm}^{-3}}
\newcommand{\countss}{\mathrm{counts}\ \mathrm{s}^{-1}}
\newcommand{\ergs}{\mathrm{erg}\ \mathrm{s}^{-1}}
\newcommand{\ergcms}{\mathrm{erg}\ \mathrm{cm}^{-2}\ \mathrm{s}^{-1}}
\newcommand{\ergcmsdeg}{\mathrm{erg}\ \mathrm{cm}^{-2}\ \mathrm{s}^{-1}~\mathrm{deg}^{-2}}
\newcommand{\ergcmssr}{\mathrm{erg}\ \mathrm{cm}^{-2}\ \mathrm{s}^{-1}\ \mathrm{sr}^{-1}}
\newcommand{\gcms}{\mathrm{g}\ \mathrm{cm}^{-2}\ \mathrm{s}^{-1}}

\newcommand{\hxdcxbcountrate}{$\sim0.01~\countss$}

\title{
Broad-band spectral analysis of the Galactic Ridge X-ray Emission
}

\author{
Takayuki Yuasa\altaffilmark{*},
Kazuo Makishima, and
Kazuhiro Nakazawa
}

\affil{
Department of Physics, School of Science,
The University of Tokyo,
7-3-1 Hongo, Bunkyo, Tokyo 113-0033, Japan
}

\altaffiltext{*}{Current affiliation and address: 
Japan Aerospace Exploration Agency (JAXA),
Institute of Space and Astronautical Science (ISAS),
3-1-1 Yoshinodai, Chuo, Sagamihara, Kanagawa 252-5210, Japan\\
yuasa@astro.isas.jaxa.jp
}

\shortauthors{T. YUASA ET AL.}
\shorttitle{BROAD-BAND SPECTRAL ANALYSIS OF THE GRXE}

\received{receipt date}
\revised{revision date}
\accepted{acceptance date}

\begin{abstract}
Detailed spectral analysis of the Galactic X-ray background emission, or the Galactic Ridge X-ray Emission (GRXE), is presented.
To study the origin of the emission, broad-band and high-quality GRXE spectra were produced from 18 pointing observations with \suzaku in the Galactic bulge region, with the total exposure of 1~Ms.
The spectra were successfully fitted by a sum of two major spectral components; a spectral model of magnetic accreting white dwarfs with a mass of $0.66^{+0.09}_{-0.07}~M_\odot$, and a softer optically-thin thermal emission with a plasma temperature of $1.2-1.5$~keV which is attributable to coronal X-ray sources.
When combined with previous studies which employed high spatial resolution of the {\it Chandra} satellite (e.g. \citealt{revnivtsevetal2009nature}),
the present spectroscopic result gives another strong support to a scenario that the GRXE is essentially an assembly of numerous discrete faint X-ray stars.
The detected GRXE flux in the hard X-ray band was used to estimate the number density of the unresolved hard X-ray sources.
When integrated over a luminosity range of $\sim10^{30}-10^{34}~\ergs$, the result is consistent with a value which was reported previously by directly resolving faint point sources.
\end{abstract}

\keywords{
X-rays: diffuse background -
Galaxy: bulge -
novae, cataclysmic variables
}

\maketitle

\section{INTRODUCTION}\label{section:introduction}\label{section:grxe_our_approach}
Apparently extended X-ray emission has been observed along the Galactic plane since early years of X-ray astrophysics (e.g. \citealt{cookeetal1969,worralletal1982ridge}), and its origin has been one of the long-standing mysteries in the research field.
After its spatial structure, the emission has been called the Galactic Ridge X-ray Emission (GRXE). Its total luminosity is estimated to be $\sim1-2\times10^{38}~\ergs$ in the conventional X-ray energy range of $2-10$~keV (e.g. \citealt{koyamaetal1986ridge67kev,valiniaetal1998ridge}).
The detection of intense emission lines from highly ionized Fe in the GRXE spectrum indicated that optically-thin thermal X-ray emission from hot plasmas must be the origin \citep{koyamaetal1986ridge67kev,yamauchiandkoyama1993}, but it was still unclear whether the plasma is gravitationally bound to discrete sources, i.e. X-ray stars, or is of truly diffuse nature filling the interstellar space. 

Many scenarios which employ different origins have been proposed to explain the GRXE. They can be divided into two major groups depending on the assumed nature of the source, namely ``Point Source" and ``Diffuse" scenarios.
The former assumes that a collection of faint but numerous discrete X-ray point sources in the Galaxy composes the GRXE (e.g. \citealt{revnivtsevetal2006}), similar to the case of the Cosmic X-ray Background (e.g. \citealt{giacconietal1979,uedaetal1999,mushotzkyetal2000}).
In the latter scenario, in contrast, literally-diffuse X-ray emitting materials are considered to fill the interstellar space over the observed scales of the GRXE, i.e.  is several tens of degrees (longitudinal) by a few degrees (latitudinal) around the Galactic center (e.g. \citealt{kanedaetal1997,tanumaetal1999} and references therein).

Unlike the ``Diffuse'' scenario which confronted several difficulties of energy supply and/or plasma confinement,
almost only one uncertainty associated to the ``Point Source'' scenario was whether there are enough number of faint Galactic X-ray sources that can collectively explain the GRXE surface brightness and spectrum, while individually satisfy the luminosity limit ($<10^{33}~\ergs$) not to be spatially resolved by X-ray telescopes.
In previous studies which used instruments with moderate angular resolutions (e.g. $>1'$), it was not possible to obtain a definitive conclusion on this problem from direct source counting.

{\it Chandra} provided a great leap in the GRXE research thanks to its ever best angular resolution in the X-ray wavelength ($\sim0.5''$).
\citet{revnivtsevetal2009nature} conducted a very deep {\it Chandra} observation on the Galactic bulge region $(l,b)=(0.113,-1.424)$, and showed that more than 80\% of the $0.5-7$~keV GRXE surface brightness can be resolved into faint X-ray stars with a limiting sensitivity of $\sim~10^{-16}~\ergcms$.
Based on X-ray spectral hardness of the faint sources, as well as their infrared identifications, \citet{ebisawaetal2005ridgeir} argued that most sources which show soft and hard spectral indices can be regarded as X-ray emitting stellar coronae and accreting white dwarf binaries like polars and intermediate polars (IPs), respectively. 
This has been also supported by a study of low-luminosity X-ray source population near the solar system \citep{sazonovetal2006luminosityfunction}.

The observation by \citet{revnivtsevetal2009nature} has invested $\sim1$~Ms of {\it Chandra} exposure in a selected field. 
This is considered to be an ultimate limit achievable with the current state of the art.
However, questions still remain: ``how do individual types of X-ray sources contribute to the GRXE?" and ``is there any additional spectral component in its spectrum that cannot be attributed to these types of sources?".

To address these issues, we take an alternative approach; broad-band and high-energy-resolution spectroscopy of the GRXE.
For understanding competitive contributions from multiple types of X-ray sources, it is essential to analyze the GRXE spectra on a broad-band basis because some of the putative constituents, namely magnetic accreting WDs, are known to emit strong signals in the hard X-ray energy band above 10~keV. High energy resolution is also necessary to understand the characteristic Fe emission lines which are emitted presumably by several types of sources.

Some authors already examined broad-band spectral similarities between the GRXE and several types of X-ray sources \citep{revnivtsevetal2006,krivonosetal2007} although spectral quality were not necessarily high. 
Compared to these reports, our study improves data quality in terms of counting statistics and energy resolution by using the \suzaku satellite \citep{mitsudaetal2007}.
In the following sections, we try to spectroscopically decompose the GRXE along  the following approach:
\begin{enumerate}[{(}i{)}]
\item Construct a broad-band GRXE spectrum using the data of \suzaku Galactic bulge observations.
\item Constrain contributions from magnetic CVs by analyzing the hard X-ray spectral shape of the GRXE using the IP spectral model constructed in \citet{yuasaetal2010} and \citet{yuasaphd}.
\item Extrapolate the magnetic CV contribution down to a softer energy band below 10~keV, and quantify contribution from other types of X-ray sources to the soft-band GRXE.
\end{enumerate}

\section{OBSERVATIONS AND PREPARATION OF DATA SET}

\subsection{The \suzaku X-ray Observatory}
{\it Suzaku}, which is the fifth Japanese X-ray astrophysical observatory launched in 2005, carries the X-ray Imaging Spectrometer (XIS; \citealt{koyamaetal2007xis}) and the Hard X-ray Detector (HXD; \citealt{takahashietal2007}).

The XIS consists of four X-ray imaging charge coupled device (CCD) cameras as a focal plane detector of the X-ray Telescope (XRT; \citealt{serlemitsosetal2007}).
The nominal energy covered by the XIS is $0.2-12$~keV \citep{koyamaetal2007xis}, and the CCD provides a rectangular field of view (FOV) of $18'\times18'$.

The HXD is a non-imaging collimated X-ray detector which consists of stacked two main detection parts; silicon $p$-intrinsic-$n$ diodes (hereafter PIN) and gadolinium silicate (GSO) scintillation crystals. They cover energy ranges of $10-70$~keV (PIN) and $40-600$~keV (GSO). In the PIN energy band, a passive collimator provides an FOV of $34'\times34'$ at full-width at half maximum, or a bottom-to-bottom aperture of $68'\times68'$ \citep{kokubunetal2007}.
This tightly collimated FOV provides one of strong points of the HXD in studies of apparently extended emission, because contamination from bright discrete X-ray sources is considerably lower than in previous hard X-ray instruments.

Thanks to the low-earth orbit (altitude of $\sim570$~km) of {\it Suzaku}, the XIS has lower non X-ray backgrounds (NXB) compared to \xmm and {\it Chandra}, and is suitable for observing extended emission with moderately low surface brightness such as the GRXE. Simultaneous and broad-band spectral coverage below and above 10~keV, achieved with the two detectors, allows us to accurately measure spectral parameters of the GRXE, including in particular the Fe-K lines and hard X-ray continua, respectively.

\begin{figure*}[htb]
\begin{center}
\includegraphics[width=0.8\hsize]{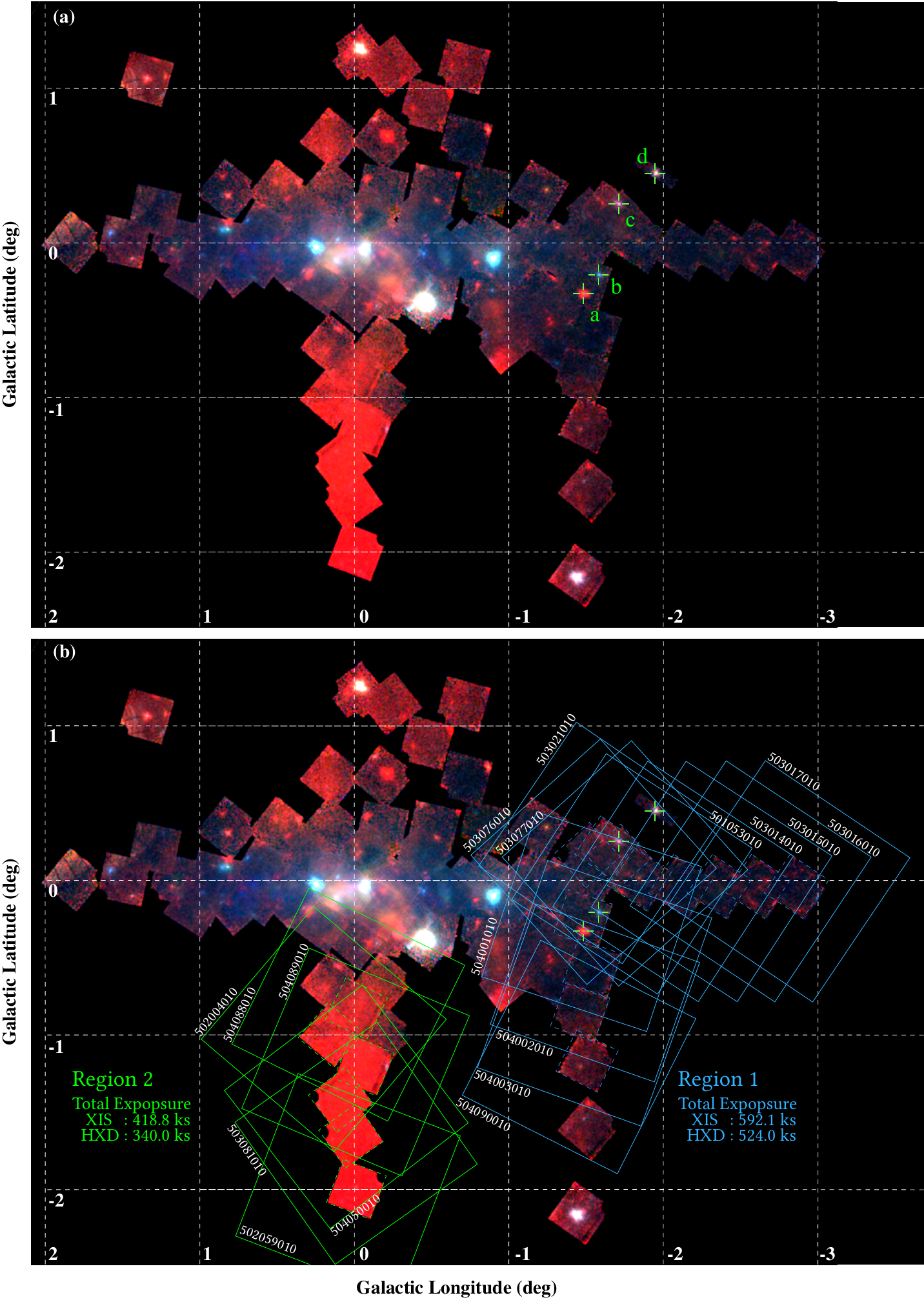}
\caption{
(a) A composite three-color X-ray image of the Galactic center and the Galactic bulge region created by adding data of 92 mapping observations performed during $2005-2010$.
It was constructed from the XIS0, 1, and 3 data. The $0.5-2$, $2-5$, and $5-8$~keV data are represented by red, green, and blue, respectively.
The image is corrected for the vignetting effect of the XRT and exposure differences, and the NXB was subtracted.
Each rectangle represents a single observation.
Labeled green crosses indicate resolved point sources of which the contributions were estimated in \S\ref{section:contami_sources}.
(b) The same image as panel a but with the fill FOVs and ObsIDs of the utilized observations (Table \ref{table:gc_observation_info}) overlaid; dashed and solid rectangles for the XIS and HXD/PIN, respectively.
}
\label{figure:gc_fovs}
\end{center}
\end{figure*}

\subsection{Mapping of the Galactic Bulge Region}
Since the launch, \suzaku has been actively performing observations of the Galactic center and the Galactic bulge regions ($|l|\lesssim5^\circ$,$|b|\lesssim3^\circ$) to study the extreme environment near the supermassive black hole, hot plasmas around supernovae, molecular clouds reflecting X-rays, and a number of X-ray point-like sources in these regions (e.g. \citealt{nobukawaetal2008,uchiyamaetalgrxe2011}).
Figure \ref{figure:gc_fovs} show a mosaic image produced by collecting all these \suzaku data with a total exposure of 4.4~Ms.
The data are taken from 92 mapping observations performed during $2005-2010$ of which individual log is summarized for reference in Table \ref{table:observation_info_gc_all}.

\subsection{Selecting Suitable Observations}\label{section:target}
To avoid signals from bright point sources which could contaminate GRXE signals, we carefully selected observations that do not contain any bright point source. 
The \integral General Reference Catalog \citep{ebisawaetal2003} and the forth {\it INTEGRAL}/IBIS catalog \citep{birdetal2010integral4thcatalogue} were used as references to positions and typical intensities of known X-ray sources. 
We discarded an observation when it is contaminated by X-ray point sources of which  flux is above 0.1~mCrab (equivalently $\sim2\times10^{-12}~\ergcms$ in the $2-8$~keV) inside the HXD FOV.
This criterion was set because a GRXE flux observed with the HXD is typically 1-2~mCrab in this sky region, and the limiting HXD/PIN flux of 0.1~mCrab corresponds to $5-10$\% of the detected GRXE flux. 
Table \ref{table:gc_observation_info} lists observations which passed the criterion, and used in the present spectral analysis. 
In Figure \ref{figure:gc_fovs} (b), the FOVs of the utilized observations are overlaid on the same X-ray image as panel (a).
To study possible spatial variations of spectral properties of the GRXE, we divided the accepted observations into two groups, and hereafter refer to them as Region 1 (on-plane, $l<0$ regions) and Region 2 (off-plane, $b<0$ regions). 
Compared to typical \suzaku observations, the total effective exposures of the selected regions are extremely deep, 592.1~ks (XIS) and 524.0~ks (HXD) for Region 1, and 418.8~ks (XIS) and 340.2~ks (HXD) for Region 2.

\begin{deluxetable*}{ccccccccc}
\tablewidth{0pt}
\tablecaption{
Summary of the Galactic bulge observations.
\label{table:gc_observation_info}
}
\tablehead{
 & \colhead{Obs. ID\tablenotemark{a}}
 & \multicolumn{2}{c}{Coordinate\tablenotemark{b}}
 & \colhead{Start time}
 & \multicolumn{2}{c}{Exposure\tablenotemark{c}}
 & \multicolumn{2}{c}{Count rate\tablenotemark{d}}\\
 &   & \colhead{$l$}  &  \colhead{$b$} &     \colhead{UT}
 & \colhead{XIS} & \colhead{HXD}   & \colhead{XIS} & \colhead{HXD}
}
\startdata
\cutinhead{Region 1}
1 & 501053010 & $-1.83$ & $-0.00$ & 2006-10-10 21:18:59 & 21.9 & 19.9 & 0.25 & 0.07\\
2 & 503014010 & $-2.10$ & $-0.05$ & 2008-09-18 04:46:49 & 55.4 & 51.2 & 0.18 & 0.05\\
3 & 503015010 & $-2.35$ & $-0.05$ & 2008-09-19 07:33:05 & 56.8 & 52.8 & 0.19 & 0.04\\
4 & 503016010 & $-2.60$ & $-0.05$ & 2008-09-22 06:47:49 & 52.2 & 49.3 & 0.18 & 0.03\\
5 & 503017010 & $-2.85$ & $-0.05$ & 2008-09-23 08:08:10 & 51.3 & 48.6 & 0.17 & 0.04\\
6 & 503021010 & $-1.62$ & $0.20$ & 2008-10-04 03:44:03 & 53.8 & 49.6 & 0.24 & 0.07\\
7 & 503076010 & $-1.50$ & $0.15$ & 2009-02-24 17:04:51 & 52.9 & 43.8 & 0.27 & 0.07\\
8 & 503077010 & $-1.70$ & $0.14$ & 2009-02-26 01:01:00 & 51.3 & 43.7 & 0.24 & 0.07\\
9 & 504001010 & $-1.47$ & $-0.26$ & 2010-02-26 09:15:00 & 51.2 & 42.2 & 0.20 & 0.05\\
10 & 504002010 & $-1.53$ & $-0.58$ & 2010-02-27 16:14:41 & 53.1 & 46.6 & 0.17 & 0.04\\
11 & 504003010 & $-1.45$ & $-0.87$ & 2010-02-25 04:33:17 & 50.9 & 41.3 & 0.19 & 0.02\\
12 & 504090010 & $-1.49$ & $-1.18$ & 2009-10-13 04:17:20 & 41.3 & 35.0 & 0.20 & 0.03\\
& & & & Total Exposure & 592.1 & 524.0 \\
\cutinhead{Region 2}
1 & 502004010 & $0.17$ & $-1.00$ & 2007-10-10 15:21:17 & 19.9 & 18.8 & 0.45 & 0.05\\
2 & 502059010 & $-0.00$ & $-2.00$ & 2007-09-29 01:40:51 & 136.8 & 110.5 & 0.35 & 0.02\\
3 & 503081010 & $0.03$ & $-1.66$ & 2009-03-09 15:41:50 & 59.2 & 57.6 & 0.49 & 0.01\\
4 & 504050010 & $0.10$ & $-1.42$ & 2010-03-06 03:55:37 & 100.4 & 80.5 & 0.60 & 0.02\\
5 & 504088010 & $-0.00$ & $-0.83$ & 2009-10-14 11:30:56 & 47.2 & 32.6 & 0.43 & 0.05\\
6 & 504089010 & $-0.05$ & $-1.20$ & 2009-10-09 04:05:59 & 55.3 & 40.2 & 0.54 & 0.02\\
&  & & & Total Exposure & 418.8 & 340.2 
\enddata
\tablenotetext{a}{Observation ID.}
\tablenotetext{b}{Aim point in the Galactic coordinate (degree).}
\tablenotetext{c}{Net exposure in units of $10^{3}$~s.}
\tablenotetext{d}{Count rates in units of counts s$^{-1}$ calculated over $1-9$ and $17-50$~keV per one of the three XIS sensors and the HXD, respectively. NXB is subtracted. The XIS count rate includes CXB which accounts about 0.04~counts~s$^{-1}$ whereas that for the HXD (\hxdcxbcountrate) is subtracted in the HXD rates.}
\end{deluxetable*}

\subsection{Contamination from Detected Point Sources}\label{section:contami_sources}

Although we selected observations with very low point-source contamination, we still notice four faint X-ray point sources, as indicated with green crosses in Figure \ref{figure:gc_fovs}, in the FOVs of Region1. Source a is a newly-found soft X-ray source, and the remaining three are previously known sources; b=AX J1742.6$-$3022 \citep{sakanoetal2002}, c=Suzaku J1740.5-3014 \citep{uchiyamaetalip2011}, and d=IGR J17391$-$3021 (e.g. \citealt{bozzoetal2010}).
When extracting the GRXE spectra from the XIS data, these sources are masked. On the other hand, we cannot exclude them in HXD/PIN because it does not have imaging capability.
As a result, the HXD/PIN spectra contain signals from these point sources, in addition to those from the GRXE.
Therefore, before actually extracting GRXE spectra, we study their effects on our HXD/PIN GRXE data.

\begin{figure*}[htb]
\begin{center}
\includegraphics[width=0.9\hsize]{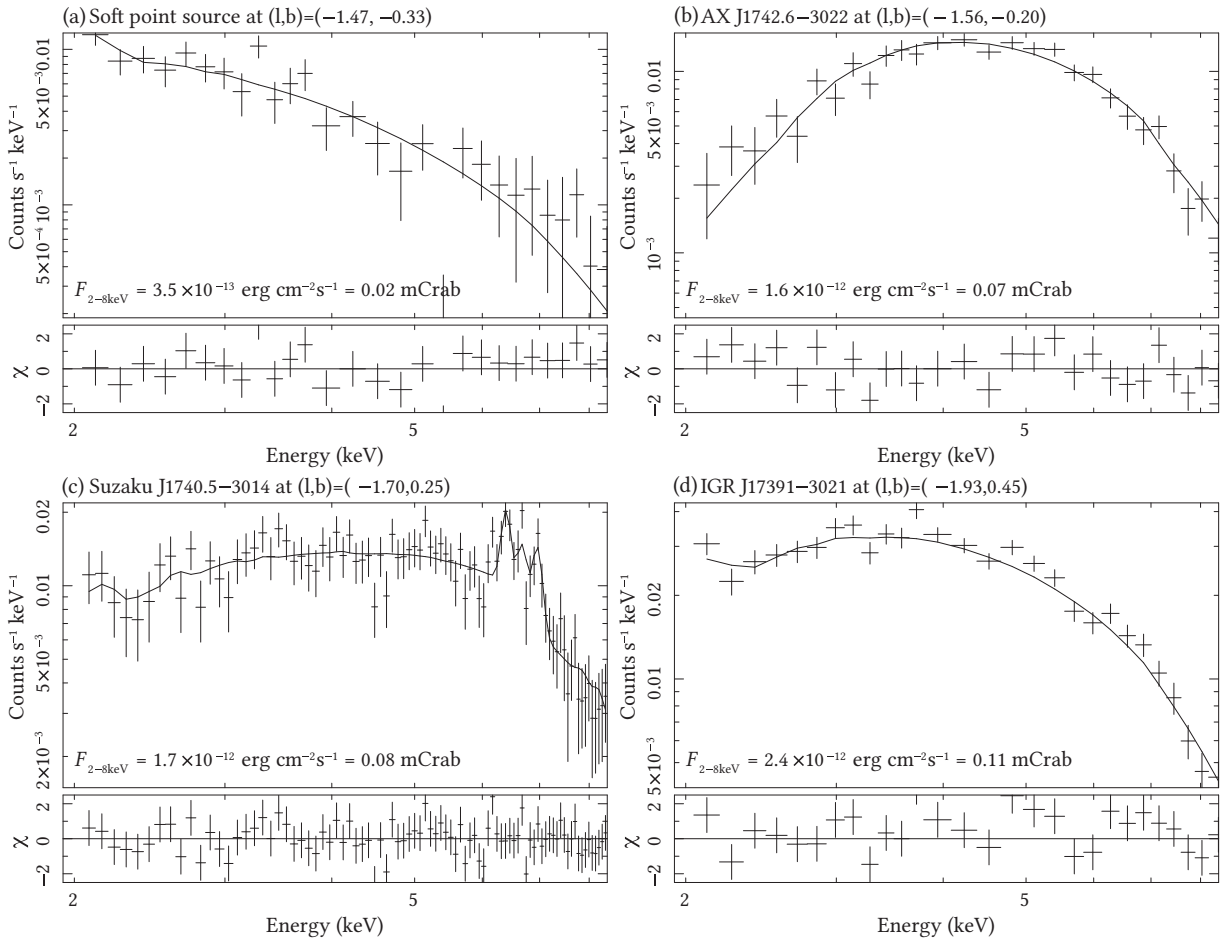}
\caption{
Background-subtracted and response-inclusive XIS spectra of the four contaminating sources recognized in Region 1; (a) a newly-found soft X-ray point source, (b) AX J1742.6$-$3022, (c) Suzaku J1740.5$-$3014, and (d) IGR J17391$-$3021.
Crosses and solid curves in upper panels are data and the best-fit models, respectively.
Lower panels present fitting residuals in terms of $\chi$. Inset labels represent model-predicted $2-8$~keV energy fluxes from the sources, followed by the same value scaled by that of the Crab nebula.
}
\label{figure:gc_contami_sources}
\end{center}
\end{figure*}

We produced spectra of the four point sources from the XIS data of Obs.IDs 504001010, 503021010, and 402066010, as shown in Figure \ref{figure:gc_contami_sources}. 
To estimate their fluxes in the HXD/PIN energy band, we fitted the spectra with phenomenological models. 
Because there is no characteristic feature in the spectra of Sources a,b, and d, we applied to them power-law models modified by the interstellar absorption (i.e. \verb|wabs|$\times$\verb|powerlaw| in the XSPEC fitting package; \citealt{arnaud1996}).
Source c shows clear emission-line features at $6-7$~keV, and is identified as an intermediate polar by \citet{uchiyamaetalip2011} based on its periodic time variation. Therefore, this source should be fitted with a multi-temperature thermal plasma model.
Because the counting statistics are not so high, we restricted ourselves to a model consisting of two collisional ionization equilibrium (CIE) plasma components plus a gaussian subject to photo absorptions [i.e. \verb|wabs|$\times($\verb|APEC|$+$\verb|Gaussian|$)+$\verb|wabs|$\times$\verb|APEC|$)$].
The two plasma components represent typical high and low temperatures of a multi-temperature plasma in an accretion column of an intermediate polar \citep[e.g.][]{yuasaetal2010}, and the Gaussian the fluorescence line from neutral Fe.

In all cases, the fits successfully reproduced the observed spectra of these sources (Figure \ref{figure:gc_contami_sources}), and provided best-fit parameters as listed in Table \ref{table:gc_contami_source}.
Model-predicted intensities in the $2-8$~keV band were $<2.4\times10^{-12}~\ergcms$ ($<0.11$~mCrab) for all the sources.
When HXD/PIN observes these sources in its FOV, the angular transmission of the collimator \citep{takahashietal2007} reduces these fluxes, and therefore their effective contributions become much smaller.

Since typical NXB-subtracted HXD/PIN count rates in Regions 1 and 2 are $\sim0.04$ and $\sim0.02~\countss$, respectively, a $\lesssim0.1$-mCrab point source, with $\lesssim0.003~\countss$, contaminates the data at most a few to ten percent of the detected counts (before reduced by the collimator angular response). We consider that these sources are negligible in the HXD band compared to the NXB subtraction uncertainty ($\sim0.003~\countss$ at 1$\sigma$).

Other than these four sources, about ten point-source-like features are seen in the XIS images.
Since their intensities, if treated as point sources, are all weak with fluxes on the order of $10^{-14}-10^{-13}~\ergcms$ in the $2-8$~keV band (equivalently $0.5-5~\mu$Crab), their total flux amounts at most to a few percent of the detected XIS count rates.
Therefore, we neglected these possible weak point sources when extracting the GRXE spectrum.

Summarizing these evaluations, we conclude that the total unwanted contribution from the resolved point sources to the HXD/PIN data is considerably less than $0.1~$mCrab, or equivalently $3\times10^{-3}~\countss$ over the $15-50$~keV band.
This means that the present analysis integrates X-ray signals emitted from unresolved sources fainter than 0.1~mCrab, or $\sim2\times10^{-12}~\ergcms$ in the $2-8$~keV band, as the ``GRXE".
This limiting flux corresponds to an intrinsic X-ray luminosity of a point source of $8\times10^{33}~\ergs$ in the same energy band if a distance to the Galactic center of 8~kpc is assumed.
This limit is several to ten times lower than those of the previous studies \citep{revnivtsevetal2006,turleretal2010} in the hard X-ray energy band, thanks to the tightly-collimated FOV of the HXD and simultaneous imaging capability of the XIS.

A further question might be raised about possible presence of bright point sources which are inside some of the HXD/PIN FOV but outside the regions covered by the XIS. However, the \integral catalog already include dim point sources which were found in the \asca Galactic center survey \citep{sakanoetal2002}, and we can conclude that there is no other bright point source which could affect our analysis. 

\begin{deluxetable*}{cccccccc}
\tablewidth{0pt}
\tablecaption{
The best-fit parameters for the contaminating point sources.
\label{table:gc_contami_source}
}
\tablehead{
 & Source & Model\tablenotemark{a} 
 & $N_{\mathrm{H}}$ & $\Gamma$\tablenotemark{b} 
 & $kT$\tablenotemark{c} & $\chi^2/\nu$ 
 & $F_{2-8~\mathrm{keV}}$\tablenotemark{d} \\
 & & & cm$^{-2}$ & & keV & & mCrab
} 
\startdata
a & New source & PL & $2.4^{+3.7}_{-1.9}$ & $3.1^{+1.5}_{-0.7}$ & -- & 1.20(25) & 0.01\\ 
b & AX J1742.6-3022 & PL & $13^{+2}_{-2}$ & $2.7^{+0.3}_{-0.4}$ & -- & 0.91(28) & 0.07\\
c & Suzaku J1740.5-3014  & 2CIE & $3.7^{+1.0}_{-1.2}$ & -- & $10.8(>6.5)$  & 0.81(86) & 0.07\\
 & & & $45^{+78}_{-35}$ & -- & $31(>4)$ & \\
d & IGR J17391-3021 & PL & $4.0^{+0.8}_{-0.6}$ & $1.9^{+0.2}_{-0.1}$ & -- & 1.32(25) & 0.11\\
\enddata
\tablenotetext{a}{PL = an absorbed power-law model. 2CIE = two collisional ionization equilibrium plasma plus a gaussian model.}
\tablenotetext{b}{Photon index of the power-law model.}
\tablenotetext{c}{Plasma temperature. Numbers in brackets are lower limits on the 90\% confidence level.}
\tablenotetext{d}{Model predicted flux in the $2-8$~keV band (1Crab$=2.2\times10^{-8}~\ergcms$).}
\end{deluxetable*}

\subsection{Time Variability}\label{section:timevariability}
If any bright point source was inside the FOV of the selected observations, and if it varied on a time scale less than $\sim$days, we should see count rate variations within individual observations.
This issue applies particularly to the HXD, because it lacks imaging capability, and outer 75\% (or 50\%, after considering angular transmission) of its FOV falls outside that of the XIS where no simultaneous imaging coverage is available.
Therefore it is also important, before performing detailed spectral analysis, to confirm that the background-subtracted signals of the HXD do not exhibit significant time variations.
This result also confirms that there was no bright fast transient point source which could contaminate the HXD data, i.e. located in the HXD FOV and outside the XIS FOV.

\begin{figure}[htb]
\centering
\includegraphics[width=\hsize]{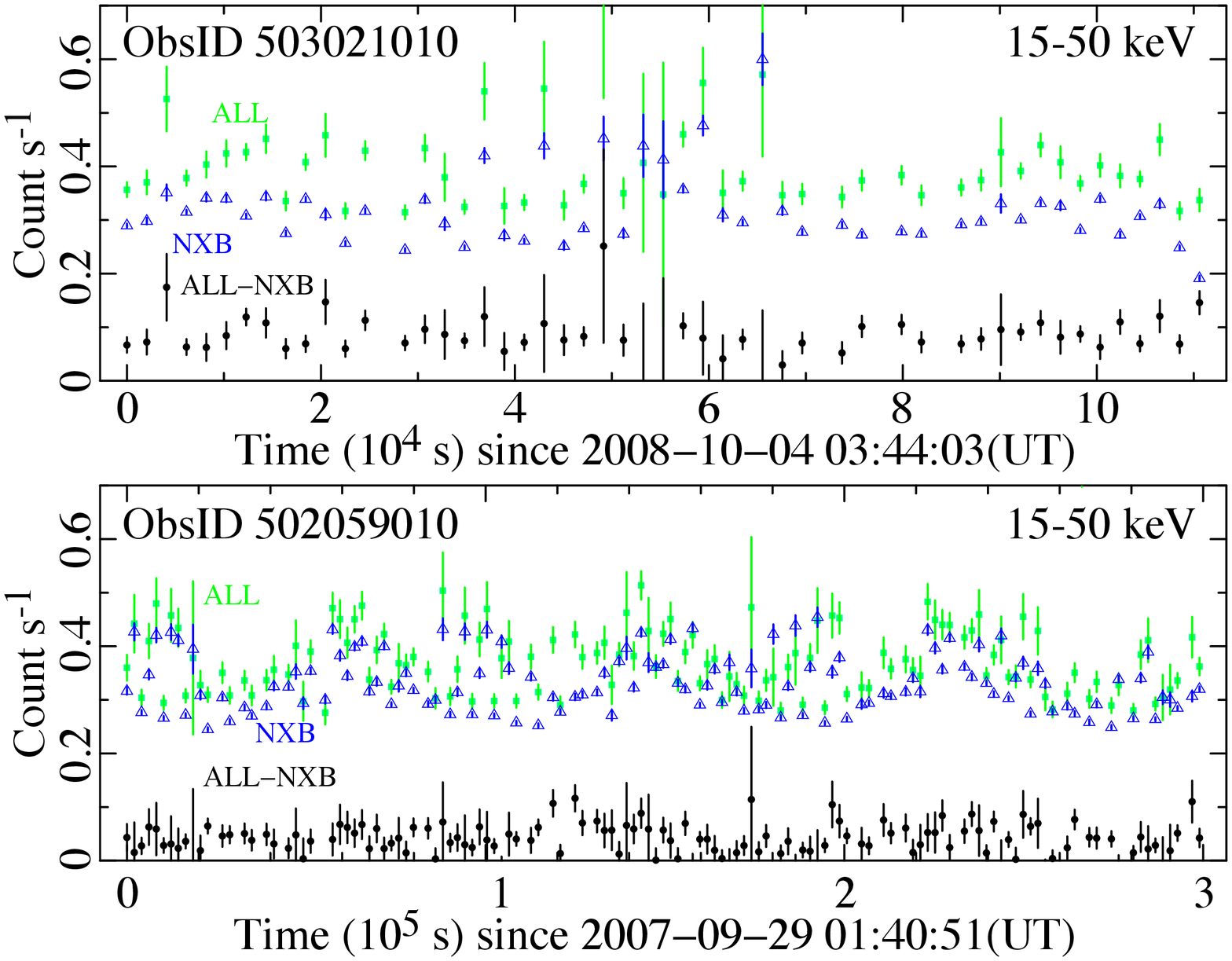}
\caption{
Examples of $15-50$~keV HXD/PIN light curves
during the GRXE observations; Obs.IDs 504002010 and 502059010 from Regions 1 and 2, respectively.
ALL (filled green rectangles), NXB (open blue triangles), and ALL$-$NXB (filled black circles) are raw detector count rates, simulated count rates of the NXB, and NXB-subtracted celestial signals including the CXB (\hxdcxbcountrate), respectively.
Each bin has a width of 2000~s.
}
\label{figure:gc_lightcurves}
\end{figure}

We extracted light curves of HXD count rates as exemplified in Figures \ref{figure:gc_lightcurves}, and examined them carefully.
The periodic variation of the total count rates (ALL in the figure) is mostly due to time variation of the NXB caused by changes of the cosmic-ray particle flux in orbit.
After subtracting the NXB, residual count rates in these observations (i.e. ALL-NXB) were $\sim0.02-0.07~\countss$, and showed little time variation. Since the CXB amounts to \hxdcxbcountrate in the HXD band, about $50-70\%$ of these residual counts can be considered to originate from the GRXE.

\subsection{Extracting the GRXE Spectrum}\label{section:gc_extracted_spectra}
Following the inspection of data quality, we extracted spectra of the individual observations in Regions 1 and 2.
The XIS data of the entire imaging area of the CCD were used, except for two corners irradiated by calibration radioactive isotopes and circularly-masked regions around the four point sources.
The XIS NXB was estimated by applying the standard tool \verb|xisnxbgen| to the same extracting region as the data.
Although the XIS detects significant signals below 2~keV as well as above that energy, we did not used them because detailed spectral modeling in this energy band is hampered by complex interstellar absorption.
As the HXD NXB, we used the simulated NXB file which is produced by the HXD team.
Accuracies of these NXB subtractions are $\sim5\%$ and 1\% ($1\sigma$) in the XIS and the HXD, respectively \citep{tawaetal2008,fukazawaetal2009}.
Since the HXD NXB amounts to $\sim0.3~\countss$ in the $15-50$~keV band, the uncertainty corresponds to $\sim0.003~\countss$. This is only $4-15\%$ of the CXB-subtracted GRXE count rates in the same energy band (Table \ref{table:gc_observation_info}).

\begin{figure}[htb]
\begin{center}
\includegraphics[width=\hsize]{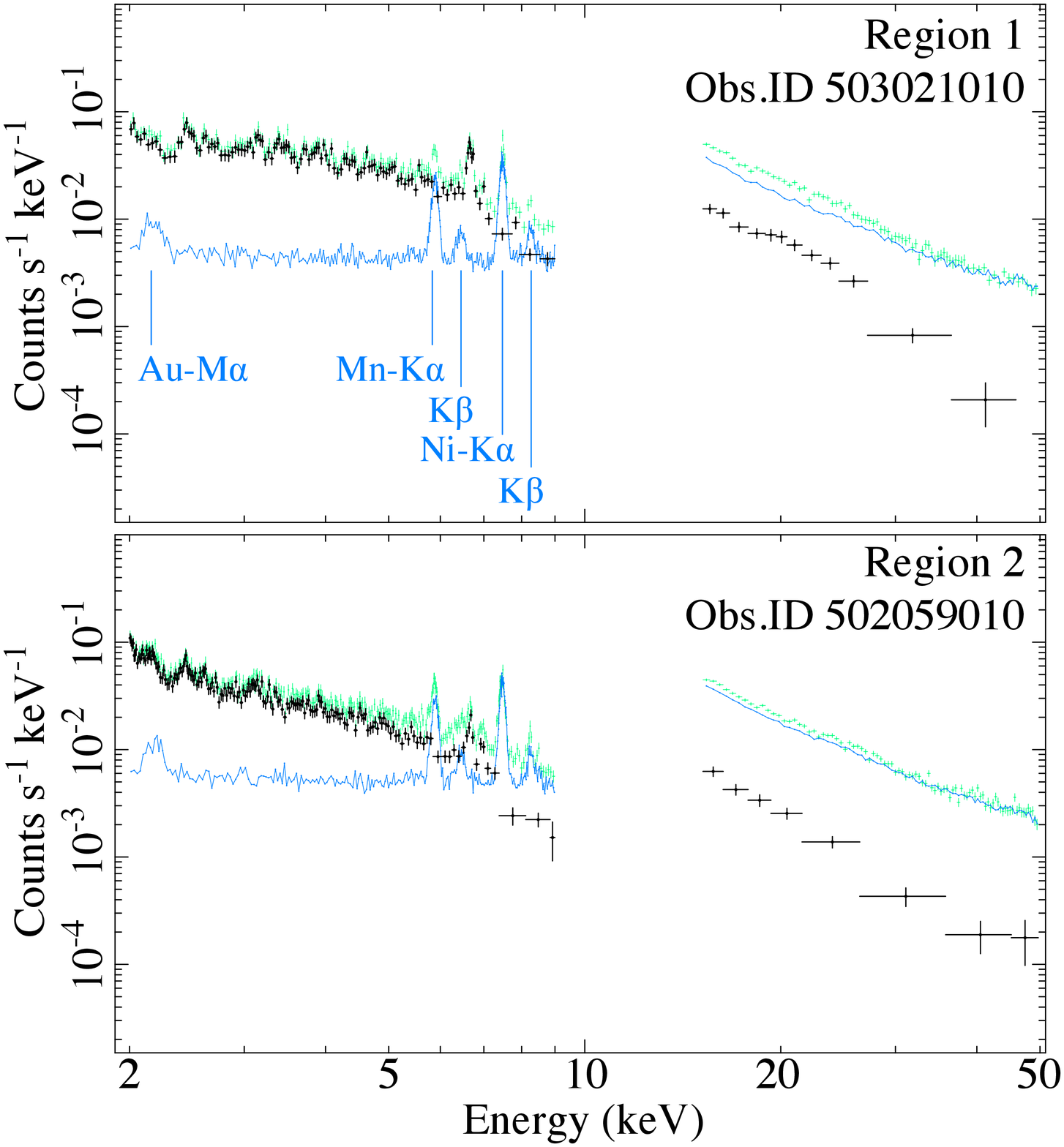}
\caption{
Examples of broad-band X-ray spectra obtained in individual observations; Obs.IDs 503021010 and 502059010 representing Regions 1 and 2, respectively.
Black crosses are the celestial signal counts, which consist of the GRXE and the CXB, derived by subtracting the NXB (blue curves) from the raw counts (green crosses). Blue labels specify atomic emission lines clearly seen in the XIS NXB (see \citealt{tawaetal2008}).
For clarity, spectra of XIS0 are plotted, although the other two XIS cameras give fully consistent data.
}
\label{figure:gc_individual_spectra}
\end{center}
\end{figure}

Figure \ref{figure:gc_individual_spectra} shows examples of the spectra, derived from the same observations as were used in the previous light curve analysis.
The raw spectra are shown as ``ALL", together with ``NXB" and ``ALL$-$NXB" (NXB-subtracted celestial signals) spectra.
As expected from the light curves, the NXB subtraction has left signals which are significant compared to the statistical and systematic uncertainties of the NXB subtraction. We regard this as the GRXE signal, again with $30-50\%$ contribution from the CXB.
A difference of spectral slopes below $\sim3-4$~keV between the two regions is caused by the stronger interstellar absorption on the Galactic plane (Region 1) than that in the off-plane regions (Region 2).
The intense Fe K$\alpha$ emission lines ($6-7$~keV) are noticeable even in the individual XIS spectra.
In addition, several emission-line like features are recognizable in $2-4$~keV (e.g. at around 2.5 and 3.2~keV) which are thought to originate from lighter elements.

\begin{figure}[htb]
\begin{center}
\includegraphics[width=\hsize]{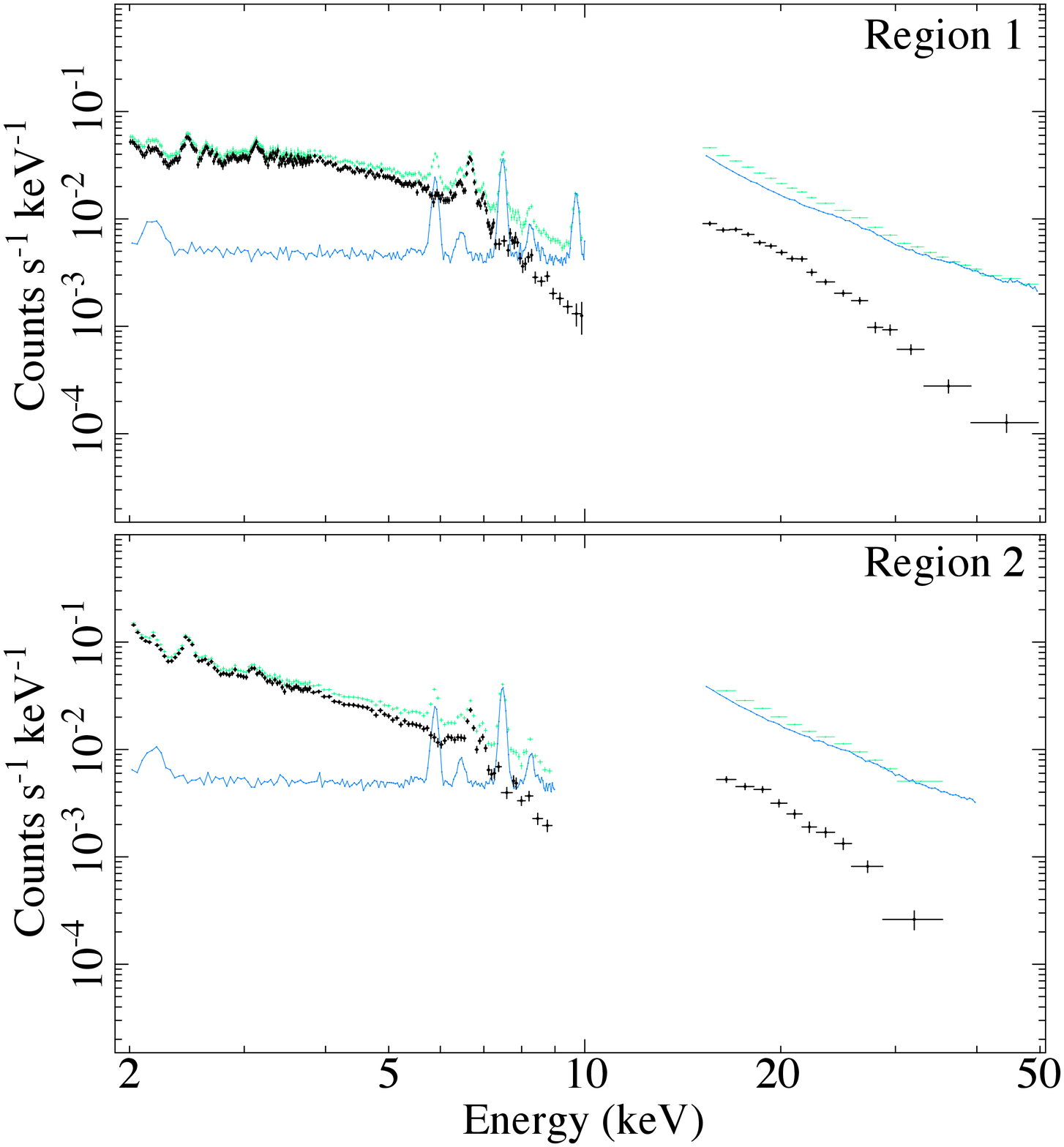}
\caption{
Spectra of the GRXE (including the CXB) summed over all observations in Region 1 (top panel) and Region 2 (bottom panel), presented in the same way as Figure \ref{figure:gc_individual_spectra}.
}
\label{figure:gc_grxe_spectra}
\end{center}
\end{figure}

\begin{figure}[htb]
\centering
\includegraphics[width=\hsize]{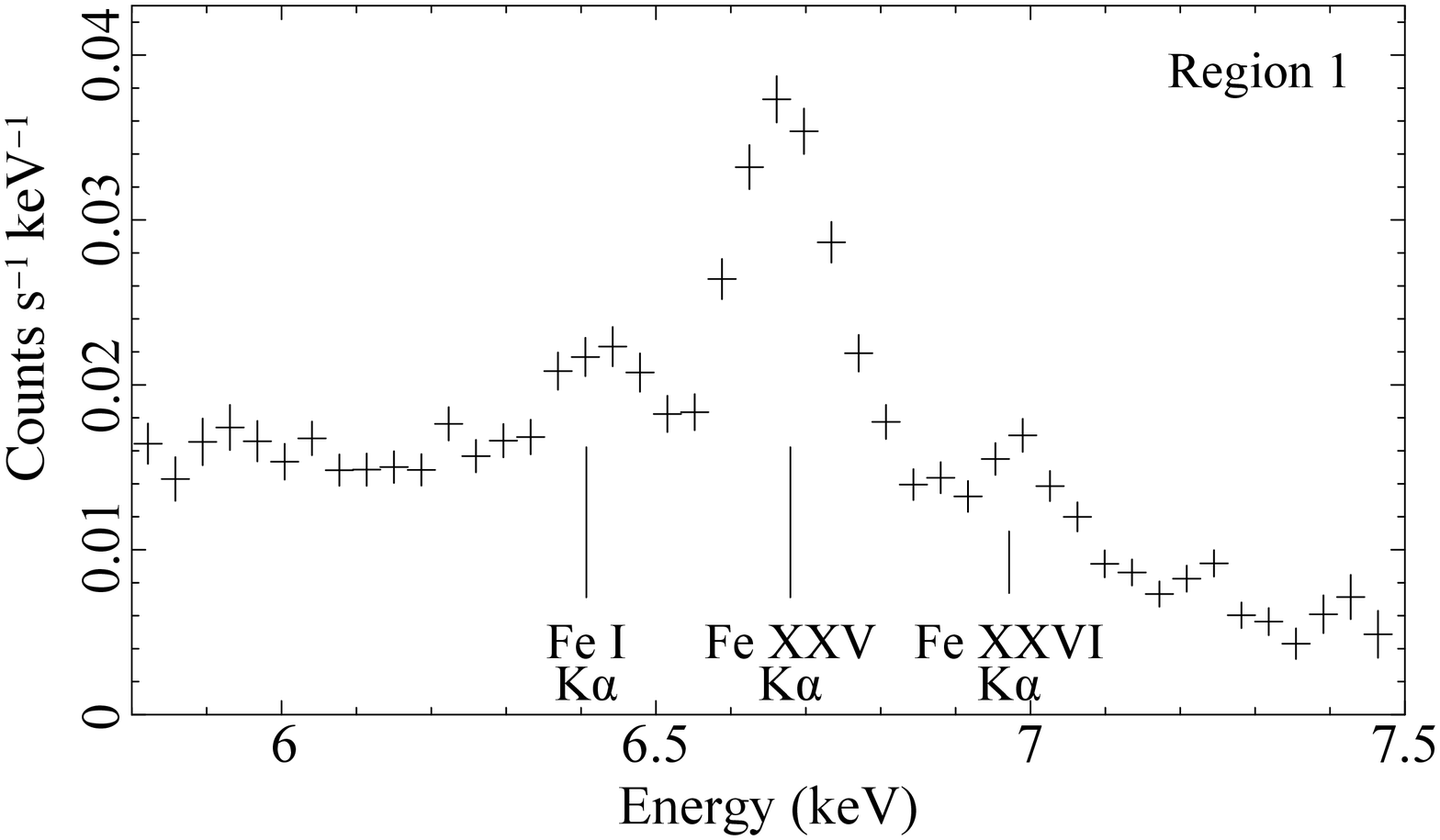}
\caption{
A close-up view of the same GRXE spectrum of Region 1 as Figure \ref{figure:gc_grxe_spectra}.
Three K$\alpha$ lines from Fe in different ionization states are clearly resolved; Fe I = neutral, Fe XXV = He-like, and Fe XXVI = H-like.
}
\label{figure:gc_grxe_spectra_fe_lines}
\end{figure}

Within each Region (1 or 2), the spectra of individual observations did not differ from one another within statistical errors.
Therefore, we combined their data for detailed broad-band spectral analyses, and created summed spectra of the two regions.
The derived spectra are shown in Figure \ref{figure:gc_grxe_spectra}.
The data summation has improved the data quality significantly, with little change in the spectral shapes.
Based on the NXB-subtraction accuracies, we consider that the celestial signals are detected up to 10~keV and 50~keV in the XIS and HXD/PIN, respectively.
Figure \ref{figure:gc_grxe_spectra_fe_lines} gives a close-up view of the Fe emission lines in the $6-7$~keV band in Figure \ref{figure:gc_grxe_spectra}.

In Region 2, we limited our spectral analysis to narrower energy ranges ($2-9$ keV for the XIS and $15-40$~keV for HXD/PIN) due to lower counting statistics, which, in turn, is because of the shorter exposure and the lower surface brightness  of the GRXE in off-plane regions.
Compared to the individual spectra, the emission lines from Fe and the other lighter elements are more clearly seen in the XIS spectra.

\subsection{Surface Brightness of the GRXE}
To conduct spectral analyses, energy responses and effective areas of the XIS and HXD/PIN should be calculated for individual observations, by taking into account a global distribution of the GRXE surface brightness.
This quantity is known to correlate with that of near infrared diffuse emission which reflects the stellar mass density \citep{revnivtsevetal2006}.
Among many plausible functions that can model the distribution, we also used, after \citet{revnivtsevetal2006}, so-called G3 model \citep{dweketal1995} for the bulge/bar structure. The model describes the stellar density $\rho(r)$ as
\begin{eqnarray}
\rho(r)&=&\rho_0 r^{-1.8} \exp(-r^3),\\
r&=&\left[ \left(\frac{x}{x_0}\right)^2 + \left(\frac{y}{y_0}\right)^2 + \left(\frac{z}{z_0}\right)^2 \right]^{1/2},
\end{eqnarray}
Where $x$, $y$ and $z$ are three-dimensional coordinates centered on the Galactic center, and $z$ axis corresponds to the pole of the Galactic rotation. See \citet{dweketal1995} for detailed explanation of the assumed coordinate system, and a rotation of the bar structure. For three scale-length parameters, $x_0$, $y_0$, and $z_0$, we used values presented in \citet{dweketal1995}; $x_0=4.01$~kpc, $y_0=1.67$~kpc, and $z_0=1.12$~kpc. 

The surface brightness $S$ at a certain sky position is assumed to follow a line-of-sight integral,
\begin{equation}
S(l,b)=S_0\int^{\infty}_0 \rho(x,y,z) \mathrm{d}s,\label{eq:nir_surface_brightness}
\end{equation}
where $(l,b)$ is a sky position in the Galactic coordinate. $S_0$ and $s$ are a normalization parameter and a line-of-sight distance to $(x,y,z)$ measured from the Sun, respectively. In this calculation, a distance between the Sun and the Galactic center was assumed to be 8.5~kpc.

\begin{figure}[htb]
\centering
\includegraphics[width=\hsize]{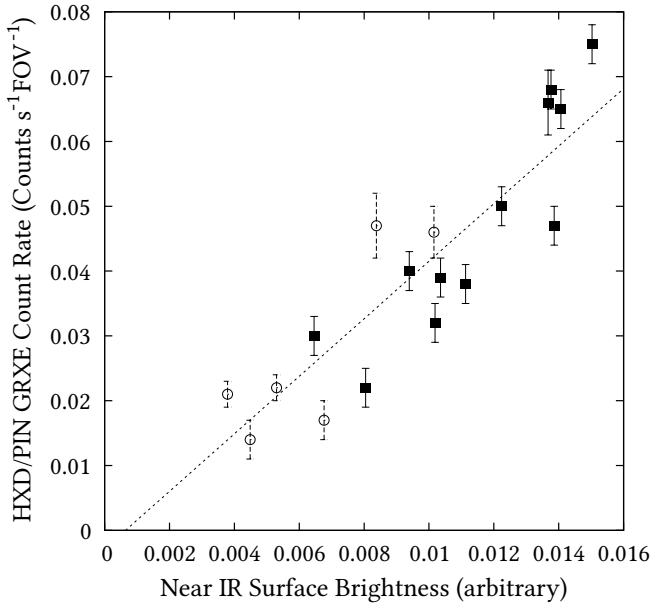}
\caption{
Correlation of the HXD/PIN GRXE count rate ($15-50$~keV) and the near IR surface brightness \citep{dweketal1995} convolved with the HXD/PIN angular transmission ($34'\times34'$ at the full width at half maximum). 
Count rates of individual observations of Regions 1 and 2 are plotted in filled rectangles and open circles, respectively.
Dashed line is the best-fit linear function ($y=4.6x-0.005$). The estimated CXB count rate of $0.016~\countss$ was subtracted from the HXD/PIN count rates.
}
\label{figure:gc_correlation_with_nir}
\end{figure}

Based on this, we constructed a surface brightness map, and used it as an input in calculating the detector responses.
To justify the procedure, we compared the actually measured HXD/PIN count rates of individual observations with the near-infrared surface brightness of Eqn. (\ref{eq:nir_surface_brightness}) convolved with the HXD/PIN angular response.
We chose HXD/PIN because the interstellar absorption is negligible unlike below 10~keV, and the CXB count rate can be accurately subtracted (whilst this is not possible in the XIS due to the absorption). A nice correlation was obtained as plotted in Figure \ref{figure:gc_correlation_with_nir}.
Thus, we, after \citet{revnivtsevetal2006}, reconfirmed that the usage of the model is appropriate in the response calculation.

\section{PARAMETERIZING THE HARD X-RAY SPECTRUM OF THE GRXE}
\subsection{Fit with the IP Spectral Model}\label{section:gc_grxe_hard_band_fit}
Magnetic CVs, including IPs in particular, are thought to be a major contributor to the GRXE in the hard X-ray band because of their relatively low luminosities, high volume densities, and the hard spectral shapes \citep{revnivtsevetal2006,sazonovetal2006luminosityfunction}. This has been directly confirmed by \chandra in the Galactic center region \citep{munoetal2004gcpointsources}.
X-ray spectral shapes of IPs and Polars are quite similar, both having the multi-temperature thermal nature; the only difference is that the plasma temperatures of Polars are lower due to enhanced cyclotron cooling (e.g. \citealt{cropperetal1998}), the GRXE spectra integrated above 15~keV should approximately be reproduced by the IP spectral model .

To test this idea, we fitted the HXD/PIN spectra of Regions 1 and 2 with the IP spectral model which we numerically constructed and verified in our previous study of nearby IPs \citep{yuasaetal2010}. This spectral model represents multi-temperature plasma emission from an accretion column on top of the magnetic poles of an IP. The WD mass and Fe abundance, which are its primary free parameters, determine the spectral shape. Compared to \citet{yuasaetal2010}, we updated the model by taking into account a plasma cooling function recently published by \citet{schureetal2009} so as to accurately model accreting gas with sub-solar abundances \citep{yuasaphd}. 
Although this affected little the spectral shape, it has increased the self-consistency of our model.

\begin{figure}[htb]
\begin{center}
\includegraphics[width=\hsize]{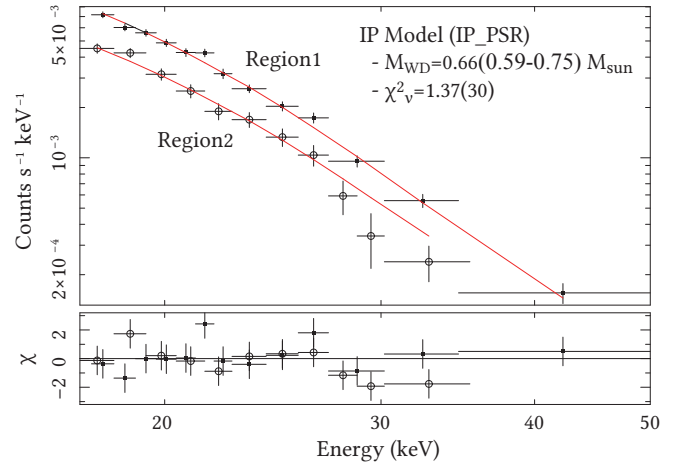}
\caption{
The HXD/PIN spectra of the GRXE of Region 1 (filled rectangles) and Region 2 (open circles), simultaneously fitted with the IP spectral model. Gray solid curves are the best-fit model spectra (CXB  inclusive). Crosses in the lower panel show fit residuals.
}
\label{figure:gc_pin_only_fit}
\end{center}
\end{figure}

In the fitting, the models for Regions 1 and 2 were constrained to share the same WD mass parameter, and allowed to have different normalizations.
The Fe abundance parameter was fixed at unity in the fitting because it cannot be constrained without data of the emission lines. Even when we changed this parameter, for example, to 0.5~solar, the result was not affected at all.
The CXB contribution is considered as a fixed model component using the established models (e.g. \citealt{boldt1987,revnivtsevetal2003}), rather than subtracting from the data, because the CXB signals are affected by interstellar absorption which is to be determined.
The best-fit model successfully reproduced the overall spectral shape in the two regions, and gave $\chi^2_\nu=1.37(30)$ with a null hypothesis probability of 9\%. Therefore, we consider this fit as acceptable. Figure \ref{figure:gc_pin_only_fit} shows the spectra, with the best-fit model superposed.
The best-fit WD mass parameter is $M_{\mathrm{WD}}=0.66^{+0.09}_{-0.07}~M_\odot$.

\subsection{Fit with simpler models}\label{section:hxd_powerlaw_fit}
To characterize the hard X-ray GRXE spectra using simpler alternatives than the IP model, we also fitted the same HXD/PIN data with a CIE plasma emission model (the \verb|APEC| model in \verb|XSPEC|) and a power-law function. Although these fittings are rather empirical, the obtained representative spectral parameters are considered to be useful for comparison with previous studies in similar energy bands.

A CIE plasma emission model yielded an acceptable fit with $\chi^2_{\nu}=1.37(31)$, and the best-fit plasma temperature of $kT=15.7^{+2.7}_{-2.0}$~keV. 
If we simply take this value as a shock temperature, and convert it to a WD mass using the relation between shock temperature and the WD mass in an IP \citep{yuasaetal2010,yuasaphd}, a WD mass of $0.48~M_\odot$ is derived.
This WD mass is lower than that obtained by the IP model fit, because the ``color temperature" of the IP model is generally lower than the actual shock temperature due to the multi-temperature nature of the post-shock region.

A power-law fit to the HXD/PIN data gave a slightly worse but acceptable fit with $\chi^2_\nu=1.43(31)$. The best-fit power-law index is $\Gamma=2.8\pm0.2$ which is slightly softer than the value derived from the \rxte and {\it CGRO}/OSSE data ($2.3\pm0.2$; \citealt{valiniaetal1998ridge}). If this power-law function is extrapolated down to the XIS energy band, the model-predicted XIS count rates exceed the actual data by more than a factor of 2 (i.e. too steep to reproduce the observed XIS spectra). Since this discrepancy will not be solved unless an artificial (and unrealistic) flattening of the power law in $\lesssim10$~keV is assumed, we consider that the power-law fit is not an appropriate model to interpret the broad-band spectrum, although it roughly reproduces the HXD/PIN spectra.

\section{SPECTRAL ANALYSIS OF THE GRXE IN THE SOFT X-RAY BAND}
\subsection{Identification of Atomic Emission Lines}
\begin{figure*}[htb]
\centering
\includegraphics[width=0.7\hsize]{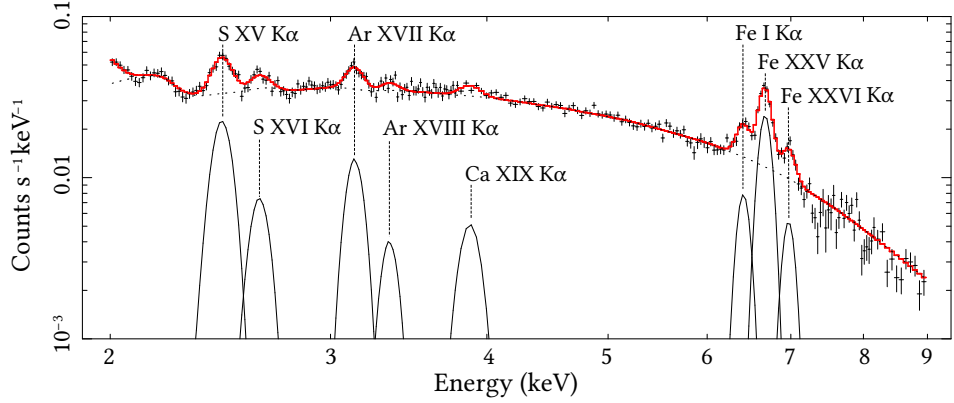}
\caption{
Emission lines seen in the GRXE spectrum of Region 1 taken with XIS0 (black crosses).
Plotted model (gray solid curve) is a phenomenological one consisting of a power-law continuum (black dotted curve) and eight Gaussians (black solid curves) for emission lines.
Only prominent lines were taken into account in the fit.
Labels represent identified emission lines.
}
\label{figure:gc_grxe_lines}
\end{figure*}

As can be seen in Figure \ref{figure:gc_grxe_spectra}, the XIS spectra exhibit many emission lines from astrophysically abundant heavy elements.
To identify elements emitting these lines, as well as ionization states, we fitted the XIS spectrum with a phenomenological model consisting of a power-law continuum subject to an interstellar absorption, and multiple Gaussians. 
As shown in Figure \ref{figure:gc_grxe_lines}, the model successfully reproduced the spectrum when we introduced 8 Gaussians for prominent line features.
The best fit center energies of these lines and their equivalent widths are listed in Table \ref{table:gc_line_identification}. With these center energies, identification is straightforward, and we found that He-like or H-like S, Ar, Ca, and Fe ions are emitting these lines as labeled in the figure.

In the $2-3$~keV energy range, the foreground diffuse X-ray emission is known to contribute to some extent to the emission lines (e.g. \citealt{kuntzsnowden2008,yoshinoetal2009,masuietal2009}).
However, the energy flux attributable to this emission is only a few percent of the detected flux of the GRXE, and therefore, is negligible at this stage.
Instead, in a broad-band spectral analysis presented below, we take into account this foreground emission.

\begin{deluxetable}{cccc}
\tablewidth{0pt}
\tablecaption{
Best-fit parameters for individual emission lines seen in the GRXE spectrum.
\label{table:gc_line_identification}
}

\tablehead{
Line & Energy & $\sigma$\tablenotemark{a} & E.W.\tablenotemark{b}\\
      &  (keV) & (eV) & (eV) 
} 
\startdata
S XV K$\alpha$ & $2.45\pm0.01$ & $<35$ & $25^{+2}_{-3}$ \\
S XVI K$\alpha$ & $2.63\pm0.01$ & $<25$ & $35^{+8}_{-9}$ \\
Ar XVII K$\alpha$ & $3.13\pm0.01$ & $<11$ & $41^{+6}_{-6}$ \\
Ar XVIII K$\alpha$ & $3.33\pm0.02$ & $<40$ & $14^{+7}_{-8}$ \\
Ca XIX K$\alpha$ & $3.86\pm0.07$ & $<89$ & $27^{+7}_{-9}$ \\
Fe I K$\alpha$ & $6.41\pm0.01$ & $<36$ & $76^{+9}_{-10}$ \\
Fe XXV K$\alpha$ & $6.68\pm0.01$ & $<17$ & $322^{+14}_{-15}$ \\
Fe XXVI K$\alpha$ & $6.97\pm0.01$ & $<34$ & $79^{+10}_{-11}$ \\
\enddata

\tablenotetext{1}{Upper limits (90\% confidence level) of intrinsic line width in terms of $1~\sigma$ of Gaussian.}
\tablenotetext{2}{Equivalent width calculated against a continuum containing the GRXE and the CXB signals.}
\end{deluxetable}

\subsection{Multi-temperature Nature of the GRXE}\label{section:gc_emission_lines}
It is almost obvious from the above identification that at least two distinct plasma components compose the GRXE.
This is because, as long as collisional ionization equilibrium is assumed, elements like S and Ar would be almost fully ionized and would not emit emission lines if the plasma is hot enough ($10^{7.5-8}$~K) to highly ionize Fe up to Fe XXVI K$\alpha$ (6.9~keV; H-like); see ionization balance calculated by, for example, \citet{bryansetal2006} and \citet{bryansetal2009}.
This reconfirms the multi-temperature nature of the GRXE first revealed by \citet{kanedaetal1997} based on similar arguments using {\it ASCA}/GIS data.
In addition, the existence of the Fe I (neutral) K$\alpha$ implies reprocessing of X-rays by a cold matter.

\begin{figure}[htb]
\begin{center}
\includegraphics[width=\hsize]{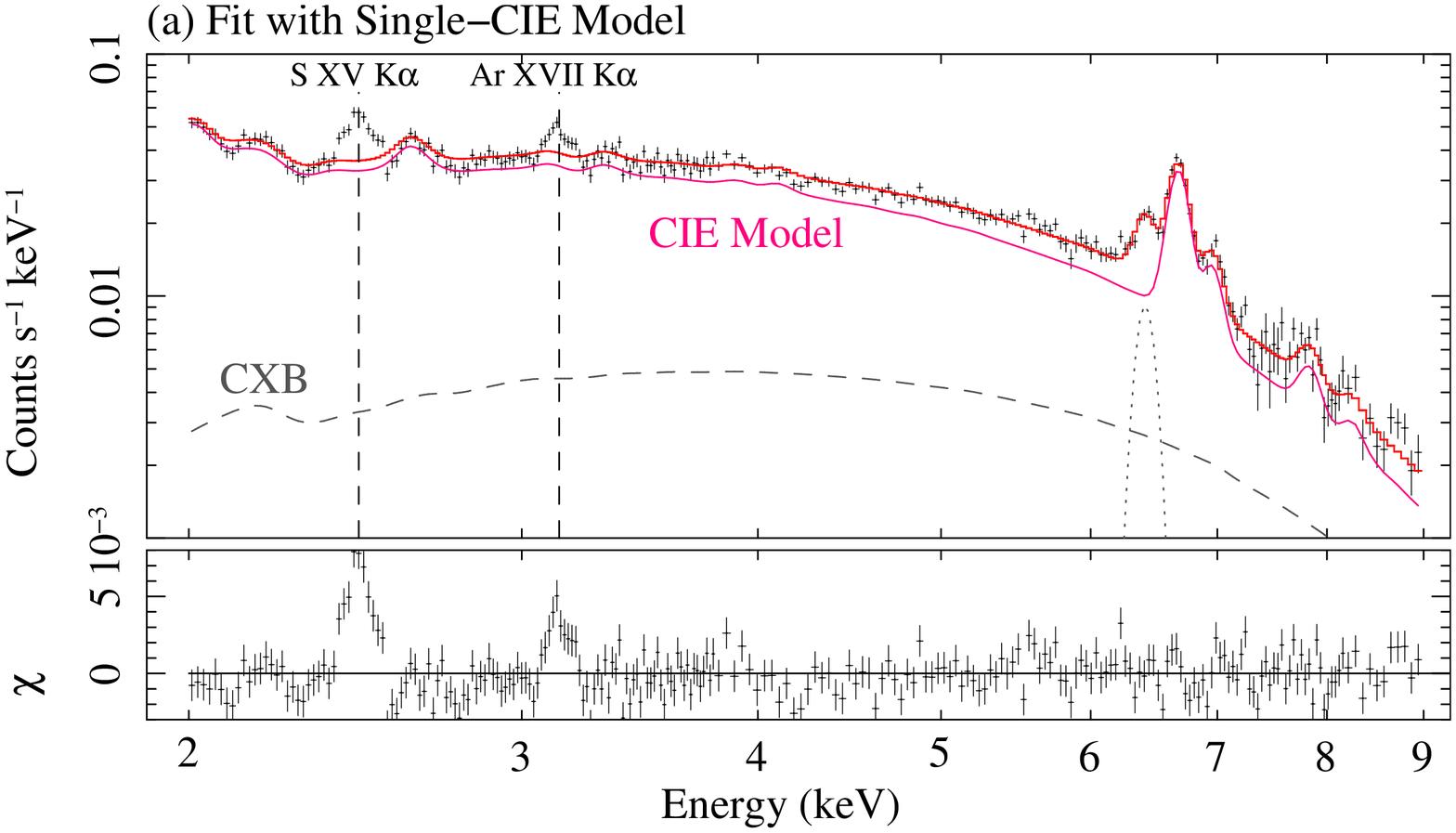}\\
\includegraphics[width=\hsize]{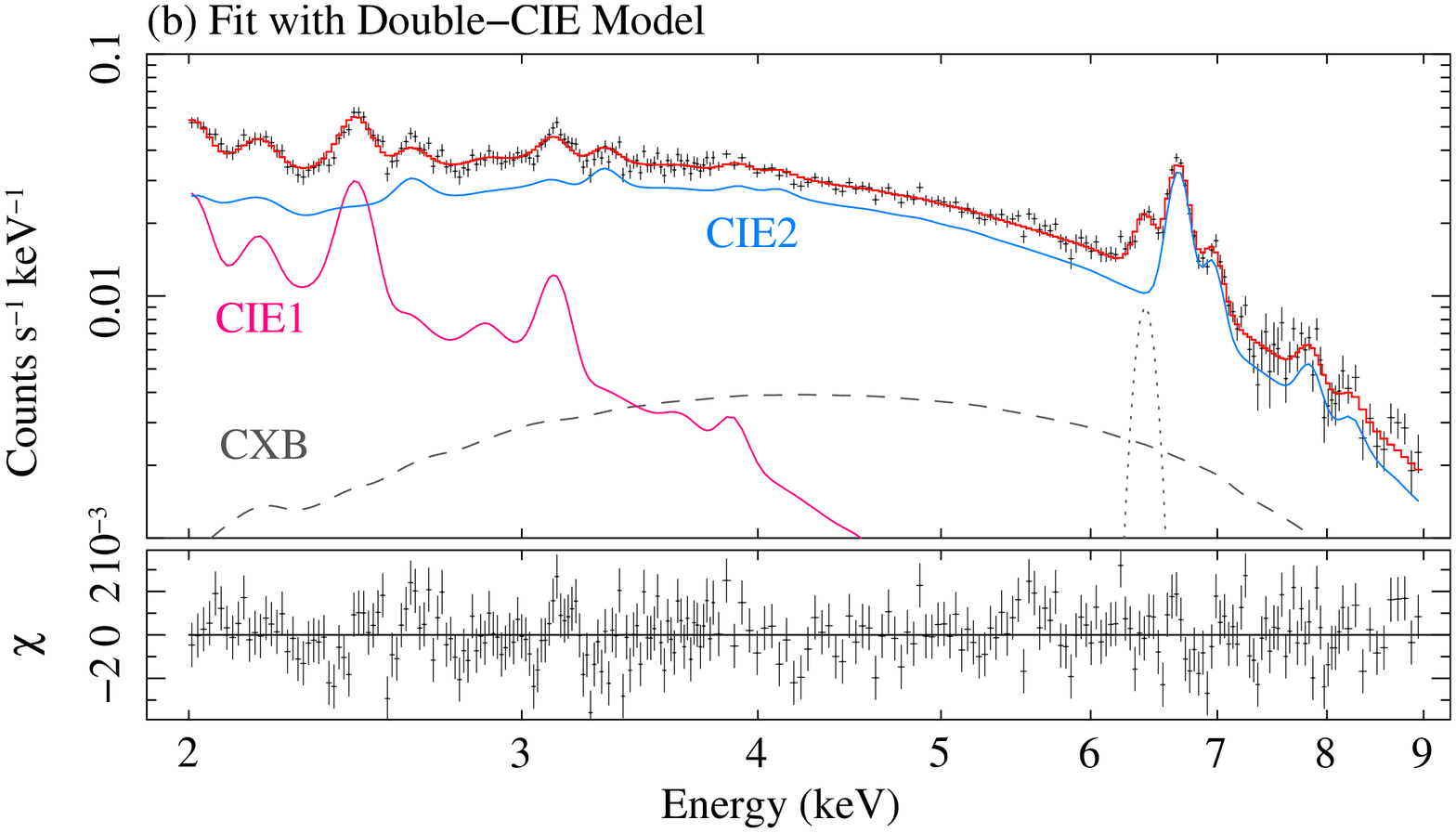}\\
\caption{
(a) The GRXE spectrum of XIS0 (black crosses) fitted with a single-temperature CIE plasma emission model and additional components, i.e. the CXB (gray dashed curve) and a Gaussian for the Fe I K$\alpha$ line (gray dotted curve).
Red and magenta curves are the total model and the CIE plasma component, respectively.
Vertical dotted lines indicate those emission lines which are not reproduced by the model.
Fitting residuals are shown in the lower panel.
(b) An improved fit incorporating two single-temperature CIE plasma emission components (magenta and blue curves). The CXB in this panel has a spectral shape which is different from that of panel a below 3~keV because the best-fit interstellar absorption column densities differ ($N_{\mathrm{H}}=2.8\times10^{22}$~cm$^{-2}$ in panel a, while $N_{\mathrm{H}}=4.8\times10^{22}$~cm$^{-2}$ in b).
}
\label{figure:gc_grxe_lines_single_apec}
\end{center}
\end{figure}

To quantify the multi-temperature nature of the GRXE found above, we first performed a fit to the same spectra with a single-temperature CIE plasma model. 
The CXB contribution was included as a fixed model, like in the HXD/PIN analysis.
As shown in Figure \ref{figure:gc_grxe_lines_single_apec} (a), the best-fit model ($kT\sim4$~keV) significantly underpredicts the S XV K$\alpha$ (2.44~keV) and Ar XVII K$\alpha$ (3.13~keV) line fluxes, and is unacceptable with $\chi^2_\nu=2.91(245)$.
These deficits cannot be resolved even when the abundance parameter is increased, because the model would then over predict He-like lines of these elements.
This reconfirms that the GRXE spectrum cannot be explained by the single-temperature CIE plasma model.

After \citet{kanedaetal1997}, we introduced an additional lower-temperature CIE component to better reproduce the low-energy lines.
The fit improved considerably to $\chi^2_\nu=1.46(243)$, and yielded the best-fit temperatures of $kT_1=0.79^{+0.46}_{-0.36}$~keV and $kT_2=6.23^{+0.26}_{-0.26}$~keV.
Although this fit, with a null hypothesis probability less than 1\%, is not yet acceptable in the strict sense, the remaining issue can be ascribed to calibration inaccuracy rather than to model inappropriateness, because the fit residuals are mainly observed in line-wing regions.
If we include a model systematic error of 2\% to simulate the insufficient response accuracy, the fit becomes formally acceptable with a null hypothesis probability of 1\%.

The derived two plasma temperatures are just representative ones, and may be approximating a mixture of more than two distinct plasma temperatures. The plasma may even have a continuous temperature distribution.
However, the data statistically do not require any additional component.
Therefore, instead of applying more complex models to the XIS data, we proceed to broad-band spectral analysis combining the XIS and the HXD/PIN data.

\section{BROAD-BAND SPECTRAL DECOMPOSITION}
As shown in \S\ref{section:gc_grxe_hard_band_fit}, the hard X-ray spectrum of the GRXE is well reproduced with spectral models which have convex spectral curvatures, or more physically, the thermal nature (i.e. the IP model or a CIE plasma model).
Besides, we reconfirmed, in the soft X-ray band, the multi-temperature nature of the GRXE \citep{kanedaetal1997}.
Based on these, we try, in this section, to reconstruct the broad-band GRXE spectrum (Figure \ref{figure:gc_grxe_spectra}) using as small a number of physically plausible spectral components as possible.

\subsection{Model Construction}
In analyzing the $2-50$~keV GRXE spectrum, let us choose, as our starting point, the two CIE plasma components which gave a successful fit to the XIS spectra.
In addition to them, as noted above, we consider a contribution from the foreground diffuse soft X-ray emission.
Based on recent reports \citep{yoshinoetal2009,masuietal2009,kimura2010}, we assumed the surface brightness of this foreground emission to be $2.5\times10^{-9}~\ergcms~\mathrm{sr}^{-1}$, and included it in the spectral model as a CIE plasma component. The temperature was fixed at $kT=0.7$~keV, because the foreground emission contributes no more than 1\%. 
Hereafter, we designate a sum of the two CIE plasma emission plus the foreground emission as ``Model 1" for simplicity.
Since the two plasma components in the GRXE can come through different absorption column densities,
they were subjected to independent absorption factors.
The abundances of major elements (i.e. Si, S, Ar, Ca, and Fe) were allowed to vary individually, but were constrained to be the same between the two CIE components.
Thus, Model 1 can be expressed as,
\begin{eqnarray}
\mathrm{Model~1}&=&\mathrm{Abs.1}\\ \nonumber 
&&\times (\mathrm{Foreground} + \mathrm{Abs.1}\times\mathrm{CXB} \nonumber\\
&&+ \mathrm{Abs.2}\times \mathrm{CIE1} + \mathrm{Abs.3}\times \mathrm{CIE2}).\nonumber
\end{eqnarray}

The 6.4-keV Fe K$\alpha$ emission line could provide a powerful tool for understanding the origin of the GRXE because, near a WD, its line shape is probably distorted by Compton scattering in the atmosphere or the gravitational redshift.
However, these studies would require a higher energy resolution.
In the present analysis, we therefore approximate it by a Gaussian with its normalization left to freely vary.

Another model, Model 2, was constructed by replacing one of the two CIE plasma components of Model 1 with our IP spectral model (\verb|IP_PSR|; \citealt{yuasaetal2010}). This can be supported by the successful application of this model to the HXD/PIN GRXE spectrum (\S\ref{section:gc_grxe_hard_band_fit}), and also by a suggestion that magnetic CVs are a major contributor to the GRXE above 10~keV (e.g. \citealt{revnivtsevetal2006}).
The Fe abundance of the IP component was tied with that of the CIE Plasma component.
Other abundance parameters were again allowed to individually vary (like Model 1).
Since intrinsic partially-covering absorption is a common feature of IPs \citep{suleimanovetal2005,yuasaetal2010}, the IP model was subjected to an additional (partially-covering) dense absorption, with a covering fraction $f$, of which absorption column density was allowed to freely vary. Summarizing above, this Model 2 is constructed as 
\begin{eqnarray*}
\mathrm{Model~2}&=&\mathrm{Abs.1}\times( \mathrm{Foreground}~+~\mathrm{Abs.1}\times\mathrm{CXB}\\
&&+~\mathrm{Fe~I~K}\alpha~+~\mathrm{Abs.2}\times\mathrm{CIE}\\
&&+~[f\cdot\mathrm{Abs.3}+(1-f)\cdot\mathrm{Abs.4}]\times\mathrm{IP\_PSR}
).
\end{eqnarray*}

We further subdivided this model into Model 2a and 2b, by treating the $M_{\mathrm{WD}}$ parameter of the IP component differently.
In Model 2a, $M_{\mathrm{WD}}$ was fixed at $0.66~M_\odot$ derived from the HXD/PIN spectral fitting (\S\ref{section:gc_grxe_hard_band_fit}). In contrast, the WD mass parameter of Model 2b was allowed to freely vary.


\subsection{Combined Spectral Fitting}\label{section:grxe_wideband_fit}
\begin{figure}[htb]
\begin{center}
\includegraphics[width=\hsize]{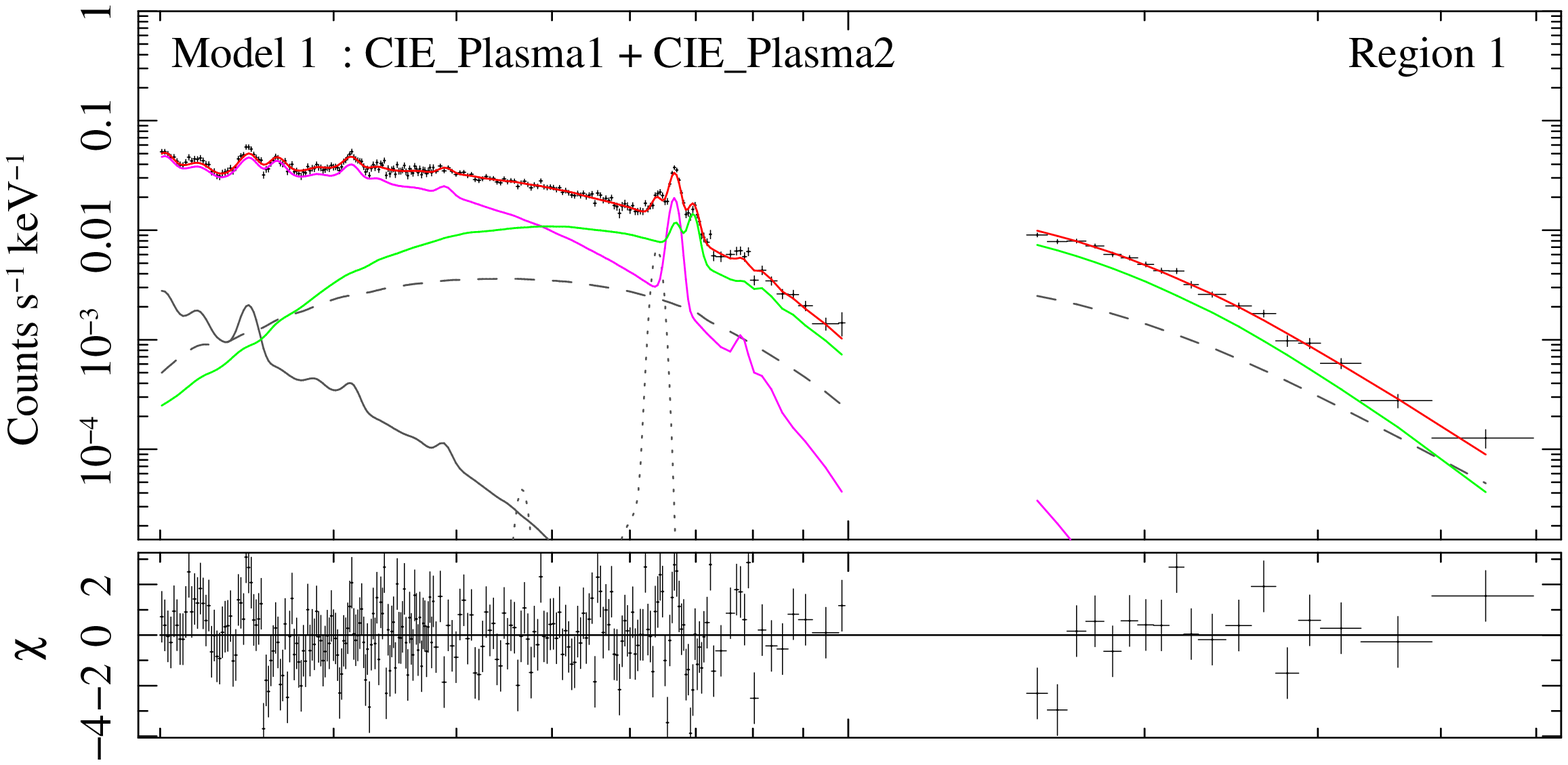}\\
\includegraphics[width=\hsize]{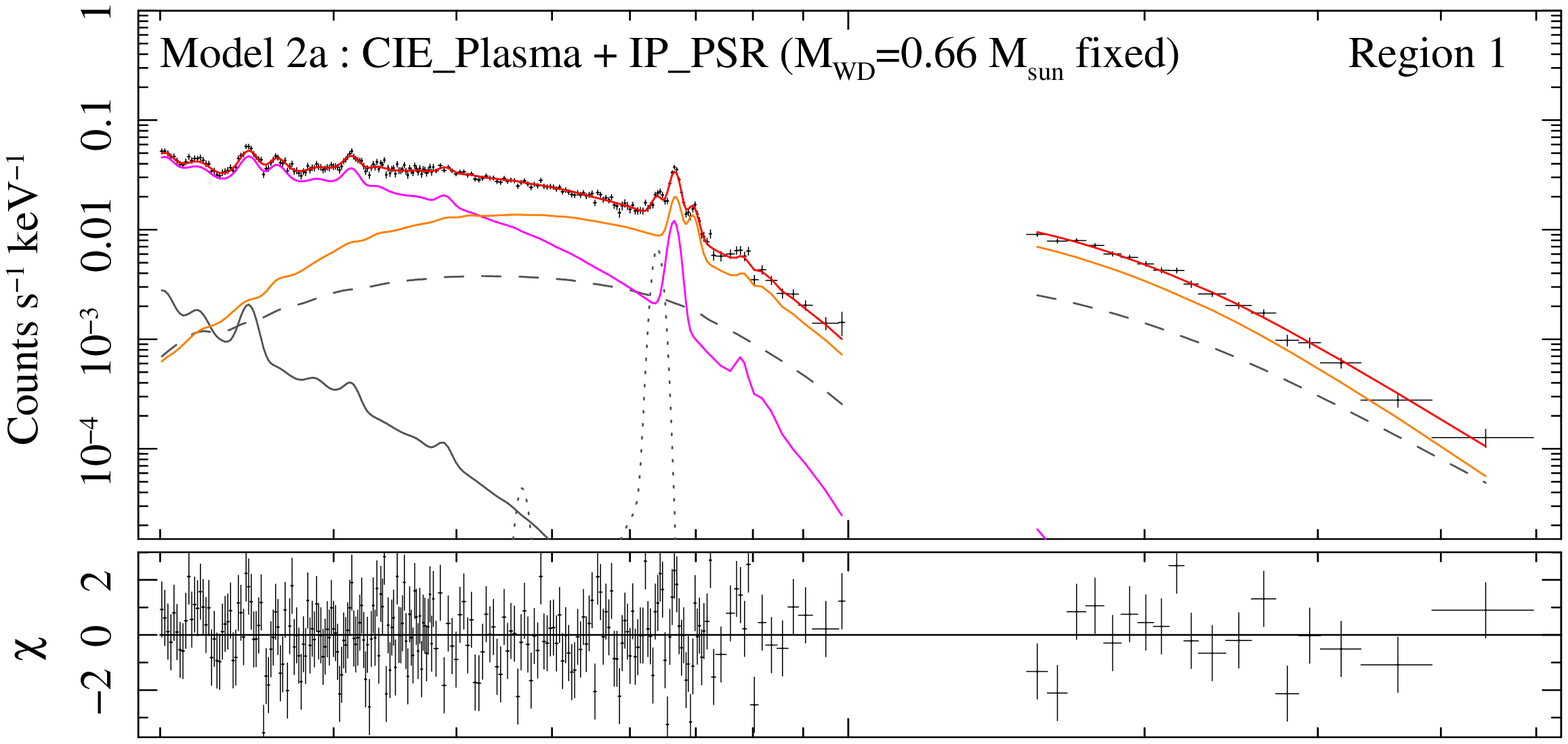}\\
\includegraphics[width=\hsize]{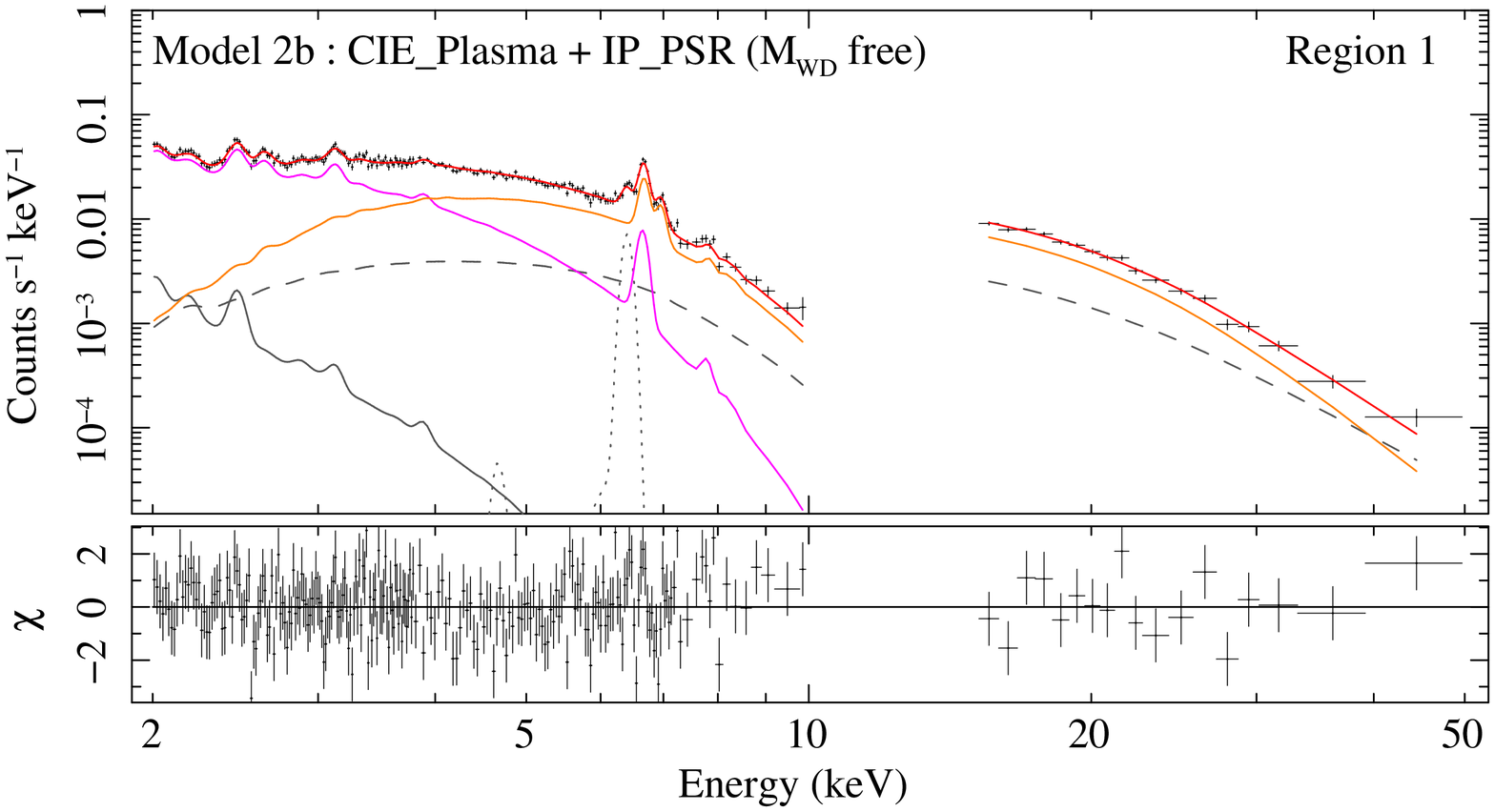}
\caption{
The GRXE spectrum of Region 1, fitted with Model 1 (top panel), Model 2a (middle panel), and Model 2b (bottom panel).
Black crosses are the observed spectrum (NXB subtracted), and red curves are the sum of all model components.
In the Model 1 fit, the two CIE plasma components are shown in magenta (lower temperature) and green (higher temperature).
Magenta and orange curves in the lower two panels are the CIE plasma and the IP model components, respectively.
Solid, dashed, and dotted gray curves are components representing the foreground diffuse soft X-ray emission, the CXB, and K$\alpha$ emission line from neutral Fe, respectively. 
}
\label{figure:gc_wideband_fit_reg1}
\end{center}
\end{figure}

We first fit the models to the stacked GRXE spectrum of Region 1, because it has higher counting statistics and covers a wider energy range over $2-50$~keV.
As drawn in Figure \ref{figure:gc_wideband_fit_reg1},
all the three models well reproduced the broad-band spectrum, and gave the best-fit parameters as listed in Tables \ref{table:gc_widebandfit_model1} and \ref{table:gc_widebandfit_model2}.
Although the values of $\chi^2$ are not yet small enough to make the fits formally acceptable, we consider again that this is caused by the inaccuracy of the XIS response in the $2-4$~keV band as explained in \S\ref{section:gc_emission_lines}.
If we include additional systematic errors (\S\ref{section:gc_emission_lines}), Model 2 became acceptable with null hypothesis probabilities larger than $1\%$, while Model 1 was still unacceptable (with a probability of $0.2\%$).


The CIE1 component of Model 1 and CIE of Model 2 yielded a plasma temperature of $kT=1.4-1.7$~keV, and account for the soft continua plus the Fe XXV K$\alpha$ line with (almost) no contribution to the Fe XXVI K$\alpha$ photons. 
Most of the hard X-ray flux detected in the HXD/PIN band is explained by the higher temperature components, CIE2 with a temperature of $15.1^{+0.4}_{-0.7}$~keV in Model 1, or the IP component with WD masses of $0.66~\msun$ and $0.48^{+0.05}_{-0.04}~\msun$ in Model 2a and 2b, respectively.

\begin{figure}[htb]
\begin{center}
\includegraphics[width=\hsize]{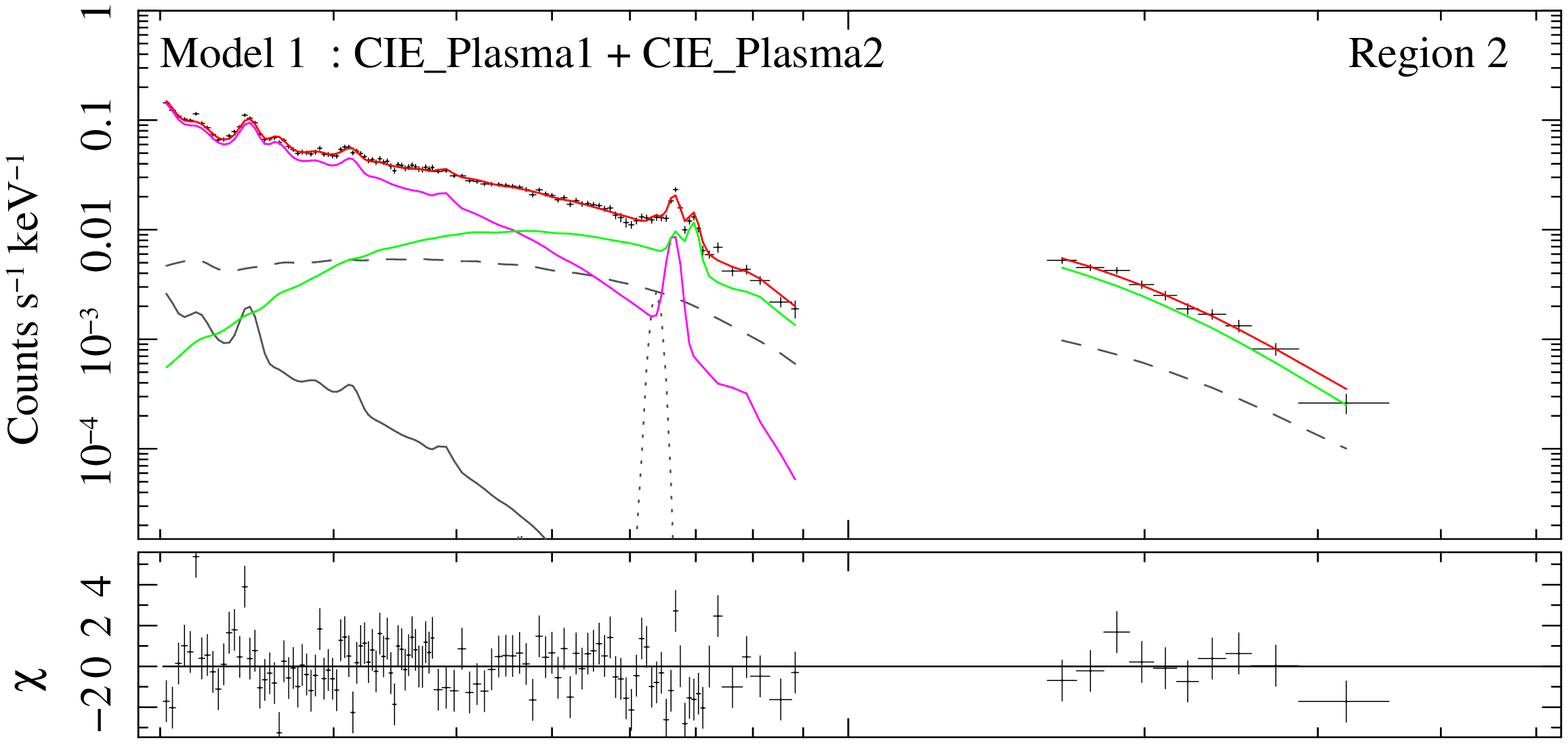}\\
\includegraphics[width=\hsize]{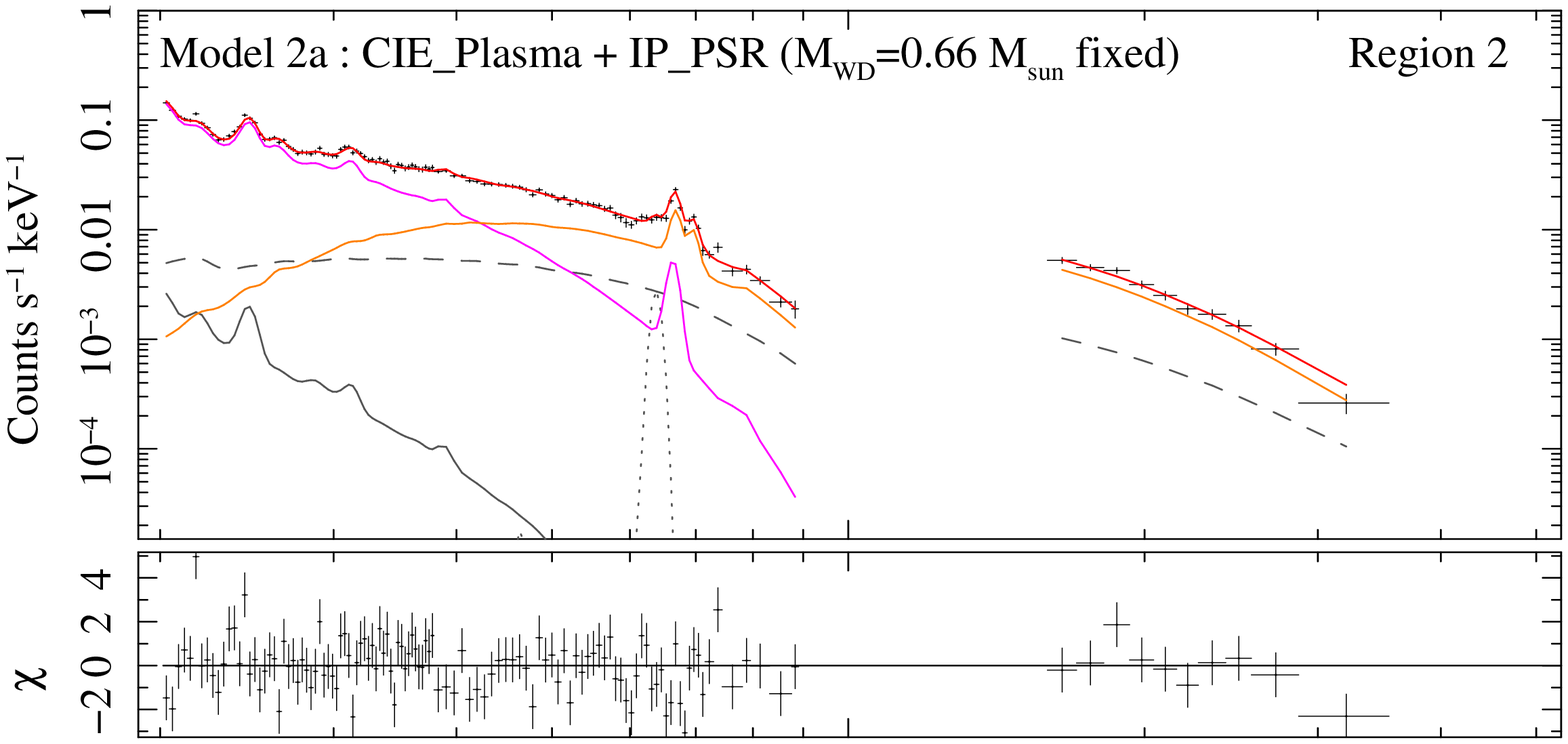}\\
\includegraphics[width=\hsize]{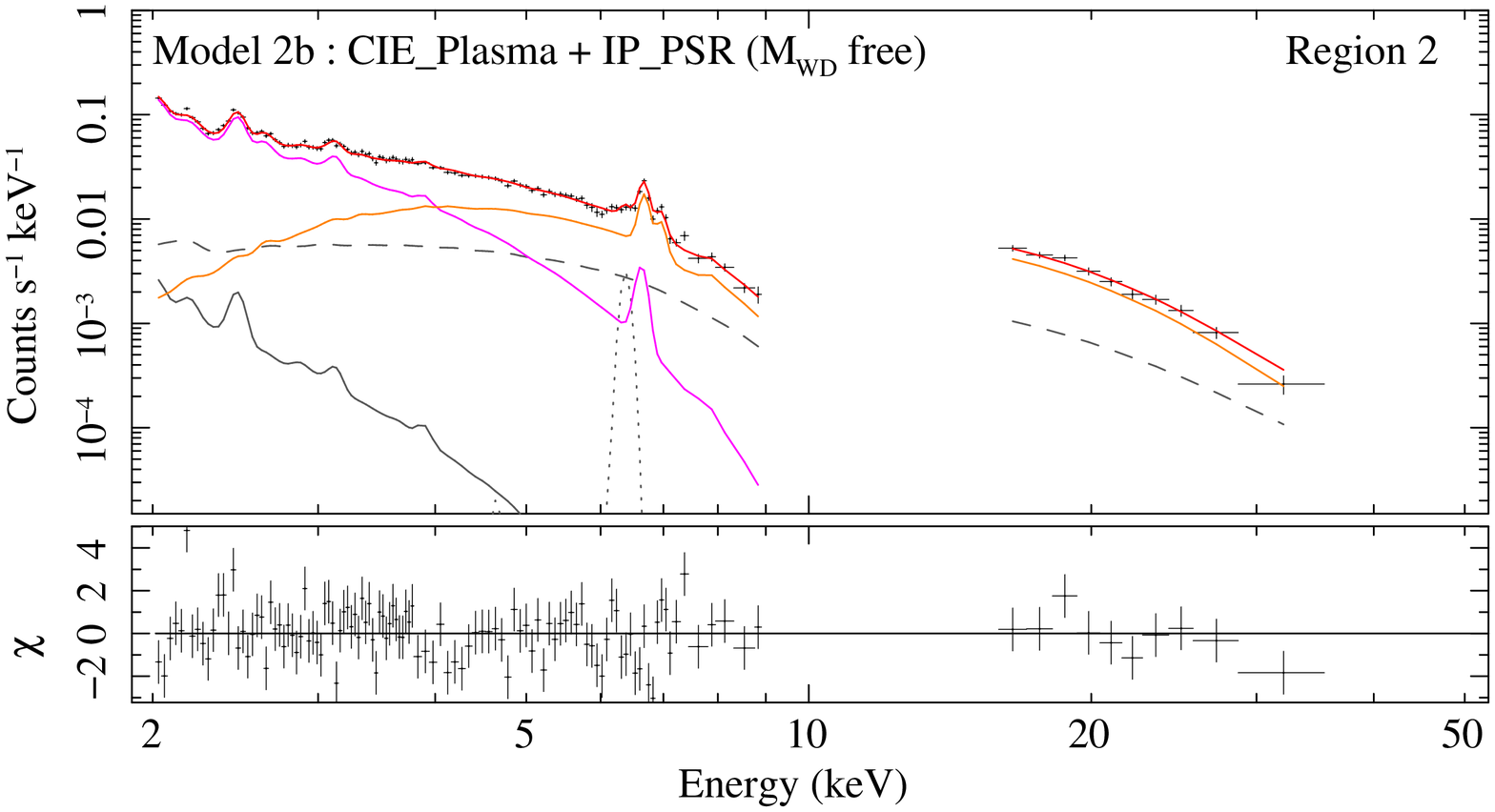}
\caption{
The same as Figure \ref{figure:gc_wideband_fit_reg1}, but for Region 2. The temperature of the lower-temperature CIE component was allowed to freely vary.}
\label{figure:gc_wideband_fit_reg2_thawed}
\end{center}
\end{figure}

Similarly, the stacked Region 2 spectrum was fitted with the same model. Due to poorer data quality, we fixed some parameters which mainly determines spectral shapes of the individual components; the CIE plasma temperatures and $M_{\mathrm{WD}}$ as summarized in Tables \ref{table:gc_widebandfit_model1} and \ref{table:gc_widebandfit_model2}. 
The best-fit models are plotted in Figure \ref{figure:gc_wideband_fit_reg2_thawed}.

\begin{figure}[htb]
\begin{center}
\includegraphics[width=\hsize]{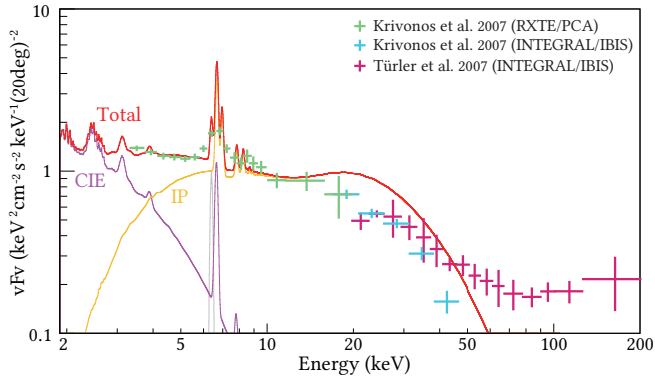}
\caption{
(a) The same best-fit GRXE model spectrum as Figure \ref{figure:gc_wideband_fit_reg1} but represented by removing the detector responses.
Data points taken from previous studies are overlaid; green and blue crosses from \citet{krivonosetal2007} and magenta crosses from \citet{turleretal2010}.
}
\label{figure:gc_model2b_vfv}
\end{center}
\end{figure}

Figure \ref{figure:gc_model2b_vfv} shows the best-fit Model 2b in Region 1 by removing the foreground emission, the CXB, and the interstellar absorption. 
A hump-like structure seen in the $10-30$~keV band is due to the partially-covering dense absorption applied to the IP model.

\begin{deluxetable*}{cccccccccccc}
\tablewidth{0pt}
\tablecaption{
Result of the wide-band spectral fitting with Model 1\tablenotemark{*}.
\label{table:gc_widebandfit_model1}
}
\tablehead{
 &
 \colhead{$n_\mathrm{{H}}$\tablenotemark{a}} &
 \colhead{$kT_\mathrm{{CIE1}}$\tablenotemark{b}} &
 \colhead{$n_\mathrm{{H}}$\tablenotemark{c}} &
 \colhead{$kT_\mathrm{{CIE2}}$\tablenotemark{b}} &
 \colhead{$Z_\mathrm{{Fe}}$\tablenotemark{d}} &
 \colhead{$\chi^2_\nu$} &
 \colhead{$\chi^2_{\nu,~3\%}$\tablenotemark{e}} &
 \colhead{$F_\mathrm{{CIE1}}$\tablenotemark{f}} &
 \colhead{$F_\mathrm{{CIE2}}$\tablenotemark{f}} \\
 &
 \colhead{$10^{22}$~cm$^{-2}$} &
 \colhead{keV} &
 \colhead{$10^{22}$~cm$^{-2}$} &
 \colhead{keV} &
 \colhead{$Z_\odot$} &
 &   
 &
 \colhead{{\footnotesize $(2-10)$}} &
 \colhead{{\footnotesize $(2-50)$}}
}
\startdata
Region 1 & $ 4.0^{+0.2}_{-0.2} $ & $1.66^{+0.04}_{-0.04} $ & $15.9^{+1.5}_{-1.3} $ & $15.1^{+0.4}_{-0.7} $ & $0.97^{+0.06}_{-0.06} $ & $1.45(754) $ & $1.15 $ & $4.9 $ & $7.8 $ \\
Region 2 & $ 1.3^{+0.2}_{-0.2} $ & $1.31^{+0.03}_{-0.03} $ & $- $ & $- $ & $1.06^{+0.07}_{-0.08} $ & $1.85(353) $ & $1.25 $ & $7.9 $ & $11.8 $
\enddata
\tablenotetext{*}{Blank parameters are fixed at the Region 1 values (see text).}
\tablenotetext{a}{Hydrogen column density of interstellar absorption.}
\tablenotetext{b}{Plasma temperatures of the CIE 1 and the CIE 2 components.}
\tablenotetext{c}{Additional absorption column density applied to the CIE 2 component.}
\tablenotetext{d}{Fe abundance of the CIE components.}
\tablenotetext{e}{Improved fitting statistics achieved when the XIS response uncertainty was virtually took into account.}
\tablenotetext{f}{Model-predicted fluxes of the CIE 1 and 2 components in $10^{-9}~\ergcms$ integrated over specified energy ranges and a $|l|<10^\circ$ and $|b|<10^\circ$ region.}
\end{deluxetable*}

\begin{deluxetable*}{ccccccccccc}
\tablewidth{0pt}
\tablecaption{
Result of the wide-band spectral fitting with Models 2a and 2b\tablenotemark{*}.
\label{table:gc_widebandfit_model2}
}
\tablehead{
 \colhead{}&
 \colhead{$n_\mathrm{{H}}$\tablenotemark{a}} &
 \colhead{$kT$} &
 \colhead{$Z_\mathrm{{Fe}}$\tablenotemark{b}} &
 \colhead{$n_\mathrm{{H}}$\tablenotemark{c}} &
 \colhead{$M_\mathrm{{WD}}$\tablenotemark{d}} &
 \colhead{$\chi^2_\nu$} &
 \colhead{$\chi^{2}_{\nu,~3\%}$\tablenotemark{e}} &
 \colhead{$F_\mathrm{{CIE}}$\tablenotemark{f}} &
 \colhead{$F_\mathrm{{IP}}$\tablenotemark{f}} & \\
 \colhead{} &
 \colhead{{\footnotesize $10^{22}$~cm$^{-2}$}} &
 \colhead{keV} &
 \colhead{$Z_\odot$} &
 \colhead{{\footnotesize $10^{22}$~cm$^{-2}$}} &
 \colhead{$M_\odot$}  &
   &
   &
 \colhead{{\footnotesize $(2-10)$}} &
 \colhead{{\footnotesize $(2-50)$}}
}
\startdata
{\footnotesize Model 2a} \\
{\footnotesize Region 1} & $ 3.6^{+0.2}_{-0.3} $ & $ 1.52^{+0.04}_{-0.04} $ & $ 0.86^{+0.05}_{-0.04} $ & $ 13.0^{+9.4}_{-1.3} $ & $ 0.66_{({\rm fixed})} $ & $ 1.30(753) $ & $ 1.04 $ & $ 3.9 $ & $8.6 $\\
{\footnotesize Region 2} & $ 1.2^{+0.2}_{-0.2} $ & $1.21^{+0.04}_{-0.03} $ & $0.82^{+0.05}_{-0.05} $ & $- $ & $- $ & $1.62(354) $ & $1.11 $ & $7.2 $ & $12.7 $\\
{\footnotesize Model 2b} \\
{\footnotesize Region 1} & $ 3.2^{+0.4}_{-0.6} $ & $1.44^{+0.06}_{-0.07} $ & $0.73^{+0.06}_{-0.05} $ & $11.8^{+8.6}_{-1.8} $ & $0.48^{+0.05}_{-0.04} $ & $1.28(752) $ & $1.02 $ & $3.2 $ & $8.6 $\\
{\footnotesize Region 2} & $ 1.1^{+0.2}_{-0.2} $ & $1.17^{+0.04}_{-0.04} $ & $0.67^{+0.04}_{-0.04} $ & $- $ & $- $ & $1.65(354) $ & $1.13 $ & $6.5 $ & $12.5 $
\enddata
\tablenotetext{*}{Blank parameters are fixed at the Region 1 values (see text).}
\tablenotetext{a}{Hydrogen column density of interstellar absorption}
\tablenotetext{b}{Fe abundance.}
\tablenotetext{c}{Additional absorption column density applied to the IP component}
\tablenotetext{d}{WD mass of the IP model.}
\tablenotetext{e}{Improved fitting statistics achieved when the XIS response uncertainty was virtually took into account.}
\tablenotetext{f}{Model-predicted fluxes of the CIE and IP components in $10^{-9}~\ergcms$ integrated over specified energy ranges and a $|l|<10^\circ$ and $|b|<10^\circ$ region.}
\end{deluxetable*}

\section{DISCUSSION}
\subsection{Interpreting the Obtained Spectral Parameters}\label{section:gc_interpret_parameters}
As presented in the previous section, the GRXE spectrum up to 50~keV can be well described only with thermal plasma models without any empirical power-law continuum. We consider that this is a strong support for our working hypothesis, the ``Point Source" scenario of the GRXE origin because the derived plasma temperatures and WD masses are well consistent with values expected from this scenario. Contrarily, in the context of the ``Diffuse'' scenario, problems of confinement and energy injection must be solved for sustain hot interstellar plasma with a temperature higher than 10~keV in the Galactic plane.

The lower-temperature CIE component (magenta in Figures \ref{figure:gc_wideband_fit_reg1} and \ref{figure:gc_wideband_fit_reg2_thawed}) resulted in ranges of $kT=1.4-1.7$~keV and $1.2-1.3$~keV in Region 1 and 2, respectively.
These temperatures are well consistent with those seen in X-ray spectra of coronal X-ray sources (e.g. \citealt{favataetal1997,covinoetal2000}) that are binary stars harboring two late-type (or ``normal") stars like the Sun, and exhibits X-ray emission via magnetic activities in their coronae.
The small temperature difference between the two regions ($\sim0.3-0.4$~keV) can be naturally understood if there is unresolved contribution from young supernova remnants (SNRs) which tend to exhibit thermal X-ray spectra of plasma (electron) temperatures of $kT\sim2-4$~keV (e.g. \citealt{kinugasatsunemi1999} for Kepler's SNR; \citealt{tamagawaetal2009} for Tycho's SNR).
Recent discoveries or possible discoveries of new SNRs in the Galactic center region \citep{sawadaetal2009,nobukawaetal2008}, especially in on-plane regions such as Region 1, support this speculation on the contribution from unresolved SNRs.

The temperature of the CIE 2 component of Model 1, $15.1^{+0.4}_{-0.7}$~keV, is consistent with a representative plasma temperature ($10-20$~keV) of magnetic CVs (e.g. \citealt{yuasaetal2010}).
The Model 2a fit successfully reproduced the broad-band spectrum although a WD mass parameter was fixed at  the value derived from the HXD/PIN analysis ($0.66^{+0.09}_{-0.07}~M_\odot$; \S\ref{section:gc_grxe_hard_band_fit}).
When it was unfixed in Model 2b, a similar but slightly lower WD mass of $0.48^{+0.05}_{-0.04}~M_\odot$ was obtained.
This reduction in WD mass is also explained by unresolved contributor from other types of CVs, especially dwarf novae in quiescent, whose spectra can also be reproduced with a CIE plasma model with lower representative temperatures ($kT\sim$a few$-10$~keV) than those of magnetic CVs.

\begin{figure}[htb]
\begin{center}
\includegraphics[width=\hsize]{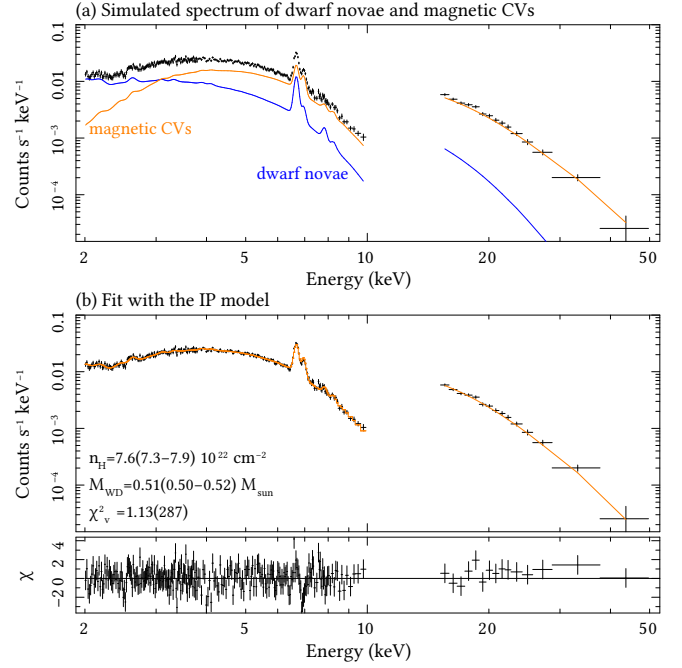}
\caption{
(a)~A simulated GRXE spectrum (black crosses) composed of spectral models of the dwarf nova (blue curve) and the magnetic CV (orange curve; the IP model).
(b)~The same spectrum as panel a, but fitted with the IP model (orange curve).
}
\label{figure:gc_simulated_dn_mcv}
\end{center}
\end{figure}

Is the reduction of the WD mass parameter by $\sim0.10-0.15~M_\odot$ realistically possible with the contribution of dwarf novae?
To test this, we simulated a composite broad-band GRXE spectrum (excluding the low-temperature CIE component) by adding spectral models of dwarf novae and IPs.
A representative plasma temperature of $kT=5$~keV and an average WD mass of $0.6~M_\odot$ were assumed for the dwarf nova and the magnetic CV components, respectively. Relative fluxes of individual components were adjusted according to X-ray emissivities of these source types measured by \citet{sazonovetal2006luminosityfunction}.
For better simulating the observed spectrum, we also applied photo absorption models with hydrogen column densities of $N_{\mathrm{H}}=3\times10^{22}$~cm$^{-2}$ (Galactic interstellar value for dwarf novae) and $10\times10^{22}$~cm$^{-2}$ (dense intrinsic absorption for magnetic CVs). Simulated count rates, or absolute intensities, were adjusted so as to match those of the observed data.
Figure \ref{figure:gc_simulated_dn_mcv} (a) presents a thus produced spectrum.
This spectrum mimics the GRXE spectrum (subtracted the low-temperature CIE component).

We performed a similar fit to this simulated spectrum using only the IP model, and obtained an acceptable fit which yields $\chi^2_\nu=1.13(287)$. The best-fit model is plotted in Figure \ref{figure:gc_simulated_dn_mcv} (b), and it 
gave the best-fitting WD mass of $0.51\pm0.01~M_\odot$ and absorption column density of $N_{\mathrm{H}}=7.6\pm0.3\times10^{22}$~cm$^{-2}$.
Thus, even when the dwarf nova component is missing in the fit model, the composite simulated spectrum can be well reproduced with the IP model with the WD mass parameter which is slightly reduced from the assumed value of $0.6~M_\odot$ ($\Delta\mwd=0.09~\msun$). This supports the above discussion on the reduced WD mass obtained in the broad-band spectral analysis.

\subsection{The Number Density of the Unresolved Hard X-ray Point Sources}\label{section:grxe_xlf_density}
The GRXE surface brightness can be used to constrain population of unresolved point sources, mostly IPs \citep{revnivtsevetal2006}.
Using the GRXE flux observed by HXD/PIN, we derive their population, or so-called X-ray luminosity function.
In the present calculation, we assume that all the point sources are located at a distance of 8~kpc from the Sun (i.e. at the Galactic center region) as done in many previous studies (e.g. \citealt{munoetal2009gcpointsources2ms}).

Generally, a luminosity function of X-ray stars $N$ can be expressed as a power-law function, and therefore we write $N$ using the intrinsic luminosity $L$
\begin{equation}
N(>L)=N_0\left(\frac{L}{L_0}\right)^{-\alpha}. \label{eqn:lognlogs_n}
\end{equation}
Here, $N(>L)$ means the surface number density of (unresolved) point sources which have an intrinsic luminosity $L$, and has a dimension of [(number) (solid angle)$^{-2}$].
$N_0$ and $L_0$ are constant factors for scaling, and the latter can be arbitrarily fixed.
The parameter $\alpha$ denotes luminosity dependence of $N$.


Now, we can calculate an X-ray surface brightness $S$ predicted from the source population $N$ by integrating 
a product of the number of sources (i.e. derivative of the cumulative density $N$) and the luminosity $L$ as
\begin{equation}
S=\int^{L_{\mathrm{min}}}_{L_{\mathrm{max}}} \frac{\mathrm{d}N(>L)}{\mathrm{d}L}~L~\mathrm{d}L \label{eqn:lognlogs_surfacebrightness}
\end{equation}
The integration is performed over a luminosity range $L_{\mathrm{min}}-L_{\mathrm{max}}$ where $\mathrm{d}N(>L)/\mathrm{d}L$ is non-zero.

For quantifying the luminosity function from the present GRXE flux,
we set $L_{\mathrm{min}}$ and $L_{\mathrm{max}}$ as follows.
The present GRXE study does not contain bright known X-ray sources which have energy fluxes higher than $\sim10^{-12}~\ergcms$ in the $2-8$~keV band, or equivalently $\sim0.1$~mCrab.
The flux limit corresponds to $1.7\times10^{-12}~\ergcms$ when expressed in the $15-50$~keV band (i.e. the HXD/PIN energy coverage). If we consider that a point source located at 8~kpc from the Sun, the flux means an intrinsic luminosity of $1.3\times10^{34}~\ergs$ in the same energy range. This should be regarded as $L_{\mathrm{max}}$.
As for $L_{\mathrm{min}}$, recent deep \chandra observations revealed that there exist X-ray sources with luminosities as low as $10^{30}~\ergs$ in the $0.5-7$~keV band \citep{revnivtsevetal2009nature}.
Another important fact is that \citet{munoetal2009gcpointsources2ms} securely measured the shape of the luminosity function down to $3\times10^{31}~\ergs$ in the $2-10$~keV band. This value provides stringent ``upper limit" for $L_{\mathrm{min}}$; i.e. its actual value should be much lower probably one order of magnitude because there is no sign of break, or turn off, in their luminosity function.
In the present study,we tentatively considered two cases with $L_{\mathrm{min}}$ being set these measured values when integrating of Eqn. (\ref{eqn:lognlogs_surfacebrightness}). If true $L_{\mathrm{min}}$ is lower, the normalization of the calculated luminosity function will decrease.

Before the integration, these luminosity values should be converted to those in the $15-50$~keV band which we concentrate on in the present calculation.
Since most of the unresolved point sources in this energy band are thought to be accreting WDs (especially magnetic ones), they have hard spectra (i.e. high plasma temperatures; \citealt{munoetal2004gcpointsources}).
We assumed, for the unresolved sources, a typical X-ray spectrum consisting of a single-temperature CIE plasma model with $kT=20$~keV suffered from the interstellar absorption of $N_{\mathrm{H}}=6\times10^{22}~\cmsq$ (typical value for the Galactic center region). Based on this spectral shape, we obtain converted $L_{\mathrm{min}}$ values of $2\times10^{30}~\ergs$ and $3.8\times10^{31}~\ergs$ in the $15-50$~keV.

\begin{figure}[htb]
\centering
\includegraphics[width=7cm]{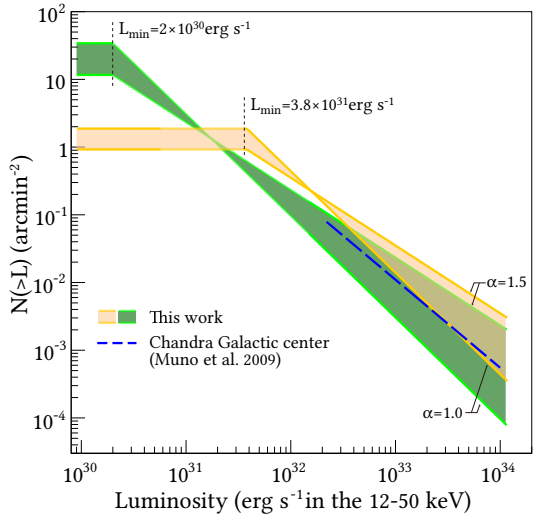}
\caption{
X-ray luminosity functions of unresolved point sources, which compose the GRXE in the higher energy band above 10~keV, calculated from the HXD/PIN GRXE flux assuming $L_{\mathrm{min}}=2\times10^{30}$ (green region) and $3.8\times10^{31}~\ergs$ (orange region) in the $15-50$~keV band.
Two solid lines which enclose each region have differently-assumed power-law indices $\alpha=1.0$ and 1.5.
Blue dashed line represents a luminosity function actually measured by {\it Chandra} in similar sky region ($2^\circ\times0.8^\circ$ around the Galactic center) in the $2-10$~keV band \citep{munoetal2009gcpointsources2ms}.
}
\label{figure:gc_lognlogs}
\end{figure}

The present HXD/PIN measurement gives the GRXE surface brightness, for example, $S=2.34\times10^{35}~\ergs~(\mathrm{PIN~FOV})^{-1}$ in Obs.ID 504001010 centered at $(l,b)=(-1.47^\circ,-0.26^\circ)$ in the $15-50$~keV band.
Since $S$, or $N(>L)$ inside it, includes two unknown parameters $\alpha$ and $N_0$, an additional constraint on one of the two parameters should be placed to determine $N(>L)$.
Previous studies \citep{ebisawaetal2005ridgeir,munoetal2009gcpointsources2ms} have revealed that the luminosity function of faint Galactic X-ray point sources has $\alpha=1.0-1.5$. Based on this, we calculate $N(>L)$ using assumed indices $\alpha=1.0$ and 1.5.

Figure \ref{figure:gc_lognlogs} shows thus calculated luminosity functions of the unresolved sources.
Two filled regions represent two representative cases with lower luminosity limits of $L_{\mathrm{min}}=2\times10^{30}~\ergs$ (green region) and $3.8\times10^{31}~\ergs$ (orange).
The luminosity functions are well consistent with that actually measured in similar sky regions by {\it Chandra} \citep{munoetal2009gcpointsources2ms}. From this comparison, we consider that the number density of required faint point sources is not unrealistically high unlike what proposed by some previous studies (e.g. \citealt{ebisawaetal2005ridgeir}).

\subsection{The Origin of the GRXE}\label{section:discussion_grxe_origin}
The broad-band spectral decomposition has been anticipated for long time for understanding the emission mechanism(s) and the energy supplier(s) of the GRXE.
This has been done in the present study, decomposing the GRXE into two representative constituents.
As was detailed in \S\ref{section:grxe_wideband_fit}, the low-temperature and the high-temperature CIE plasma emissions have plasma temperatures of $\sim1$~keV and $>10$~keV according to our modeling. The latter component can be successfully replaced by the IP spectral model as examined in the Model 2 fits.
In addition, the luminosity function of unresolved hard X-ray sources does not largely contradict with one directly measured with {\it Chandra}.
Based on these, we consider that the present results in $<10$~keV provide yet another very strong support for the ``Point Source" scenario of the GRXE alongside of the imaging decomposition of \citet{revnivtsevetal2009nature}.

On the other hand, the results are against for several previous studies that suggested the ``Diffuse" scenario as the GRXE origin.
The putative non-equilibrium ionization (NEI) plasma proposed by \citet{kanedaetal1997} is rejected since we detected intense Fe XXVI K$\alpha$ line in the present sky region which is not expected from their NEI model.
This is a confirmation of a similar denial argument by \citet{ebisawaetal2008} in the different sky region.
Unlike \citet{yamasakietal1997ridge} and \citet{valiniaetal1998ridge}, the observed broad-band spectrum did not require a putative hard-tail component which was suggested to smoothly connect to the gamma-ray background emission up to hundreds of MeV.

\subsection{The Mean WD Mass in the Galaxy}
The WD mass of CVs is an important parameter when interpreting the hard X-ray GRXE spectral shape.
It was first estimated to be $\sim0.5~M_\odot$ in \citet{krivonosetal2007} by roughly fitting their GRXE spectra with the IP model by \citet{suleimanovetal2005}.
Our hard X-ray analysis confirmed this, and gave a slightly heavier WD mass of $0.66^{+0.09}_{-0.07}~M_\odot$.
As \citet{krivonosetal2007} mentioned, the value could be interpreted as the mean WD mass of cumulated magnetic CVs in the Galaxy.

If we accept this idea, it is intriguing to compare the derived WD mass with that reported for isolated WDs based on optical spectroscopy. For example, using large SDSS data, \citet{kepleretal2007} reported a mean WD mass of $0.593\pm0.016~M_\odot$ for 1733 WDs which have helium and hydrogen outer layers.
Our value $0.66^{+0.09}_{-0.07}~M_\odot$ includes the average within errors.
The slight difference of the center values ($\sim0.06~\msun$) could be a result of long-lasting mass accretion.
However, CVs are thought to evolve through the common-envelope phase of main-sequence stars, in which two stars in a binary share their outer layers, possibly resulting in WD masses different from those of isolated stars.
Therefore, at this moment, it is difficult to investigate the difference of the average WD masses.
More accurate WD mass determinations in CVs, and sophisticated evolution model of them are necessary to accurately perform above comparison.

\subsection{Connection to the Galactic Center X-ray Emission}
A low-temperature CIE component ($kT\sim1$~keV) similar to one seen in the GRXE was also reported recently by \citet{nobukawaetal2010} in the Galactic center X-ray emission observed in $|l|<0.2^\circ$. 
Probably due to lack of the hard X-ray spectral coverage in their analysis, the authors reported a plasma temperature of the hotter CIE component of $7.0\pm0.1$~keV, which is considerably cooler than our result $kT=15.1^{+0.4}_{-0.7}$~keV.
Note also that their spectral analysis introduced, in addition to two CIE plasma components, an intense power-law component which was not detected in our broad-band GRXE data.

In the Galactic center region, the super-massive black hole ($\sim3\times10^6~M_\odot$) is suspected to have a close connection with the unresolved (or literally diffuse) emission which is observed in its proximity (e.g. predicted by \citealt{sunyaevetal1993} and observed e.g. by \citealt{inuietal2009}). This emission could have another origin which differ from that of the GRXE.
Besides, stellar density is higher by more than an order of magnitude in the very central region of the Galaxy ($\sim100$~pc from the Galactic center) compared to those of the region we studied.
Therefore, X-rays from faint point sources (mostly coronal X-ray sources and IPs) might occupy a considerable fraction of the detected signals, although \citet{nobukawaetal2010} simply neglected them.
Since about a half of such faint point sources radiate its energy in the hard X-ray band,
we stress that a broad-band spectral analysis including the hard X-ray energy range is essentially powerful to examine possible existence of truly diffuse emission in the Galactic center.

\section{SUMMARY}
We accumulated the data of \suzaku GRXE observations in the Galactic bulge region
achieving $\sim1$~Ms exposure in total (\S\ref{section:target}), and produced the broad-band GRXE spectra with high counting statistics covering the $2-50$~keV band (\S\ref{section:gc_extracted_spectra}).
We showed that the hard X-ray GRXE spectrum taken with HXD/PIN is well reproduced by the IP spectral model (\S\ref{section:gc_grxe_hard_band_fit}) with a WD mass parameter of $0.66^{+0.09}_{-0.07}~M_\odot$.
From emission line analyses based on the CIE modeling which is more appropriate than previous line studies (e.g. \citealt{kanedaetal1997}), we reconfirmed the multi-temperature nature of the GRXE (\S\ref{section:gc_emission_lines}).

Based on these results, we constructed physical models of the GRXE (\S\ref{section:grxe_wideband_fit}) which consists of multi-temperature CIE plasma emissions. The models nicely decomposed the broad-band GRXE spectral shape for the first time. Especially, the model that includes the IP component, namely Model 2, gave better fits to the data together with the low-temperature CIE plasma component.
The derived plasma temperature and the WD mass are quite consistent with those of coronal X-ray sources and typical of magnetic accreting WDs (\S\ref{section:gc_interpret_parameters}).

We also calculated the X-ray luminosity function and the number density of the unresolved hard X-ray point sources, mostly IPs (\S\ref{section:grxe_xlf_density}).
The luminosity function is consistent with that obtained from a deep imaging observation near our field, indicating that no unknown X-ray source type is additionally required to explain the GRXE flux.

Combining the spectral decomposition result and the calculated source density,
we concluded that the present result supports the ``Point Source" scenario of the GRXE origin.

\appendix
\section{{\it SUZAKU} OBSERVATIONS IN THE GALACTIC CENTER REGION}\label{section:gc_all}
For later reference, we summarize information of 92 {\it Suzaku} observations in the Galactic center region in Table \ref{table:observation_info_gc_all}. XIS data of these observations are used to produce the mosaic image shown in Figure \ref{figure:gc_fovs}.
The total effective exposures of the XIS and HXD/PIN are 4.4 and 3.9~Ms.
The list contains observations with various types of scientific aims; the Galactic Ridge X-ray Emission, supernova remnants, transient X-ray binaries, molecular clouds reflecting X-rays, and unidentified sources in TeV wavelength (H.E.S.S unID sources).

\begin{deluxetable}{cccccrr}
\tablewidth{0pt}
\tablecaption{
\it Suzaku \rm observations of the Galactic center.
\label{table:observation_info_gc_all}
}

\tabletypesize{\scriptsize}

\tablehead{
 &
 \colhead{Obs. ID\tablenotemark{a}} &
 \multicolumn{2}{c}{Coordinate\tablenotemark{b}} &
 \colhead{Start time} &
 \multicolumn{2}{c}{Exposure\tablenotemark{c}} \\
 & 
 \colhead{} &
 \colhead{$l$} &
 \colhead{$b$} &
 \colhead{UT} &
 \colhead{XIS} &
 \colhead{PIN}
}
\startdata
1 & 100027010 & $0.06$ & $-0.08$ & 2005-09-23 07:18:25 & 44.8 & 37.9 \\
2 & 100027020 & $-0.24$ & $-0.05$ & 2005-09-24 14:17:17 & 42.8 & 36.1 \\
3 & 100027030 & $-0.44$ & $-0.39$ & 2005-09-24 11:07:08 & 2.1 & 1.9 \\
4 & 100027040 & $-0.44$ & $-0.07$ & 2005-09-24 12:41:33 & 1.9 & 1.8 \\
5 & 100027050 & $0.33$ & $0.01$ & 2005-09-25 17:29:12 & 2.0 & 1.8 \\
6 & 100037010 & $-0.24$ & $-0.05$ & 2005-09-29 04:35:41 & 43.7 & 39.4 \\
7 & 100037020 & $-0.44$ & $-0.39$ & 2005-09-30 04:30:44 & 3.3 & 3.1 \\
8 & 100037030 & $-0.45$ & $-0.07$ & 2005-09-30 06:06:32 & 3.0 & 2.8 \\
9 & 100037040 & $0.06$ & $-0.08$ & 2005-09-30 07:43:01 & 43.0 & 39.5 \\
10 & 100037050 & $0.33$ & $0.01$ & 2005-10-01 06:22:41 & 2.4 & 2.2 \\
11 & 100037060 & $0.64$ & $-0.10$ & 2005-10-10 12:28:01 & 76.6 & 70.8 \\
12 & 100037070 & $1.00$ & $-0.10$ & 2005-10-12 07:10:24 & 9.2 & 9.5 \\
13 & 100048010 & $0.06$ & $-0.08$ & 2006-09-08 02:23:24 & 63.0 & 60.3 \\
14 & 102013010 & $0.06$ & $-0.08$ & 2007-09-03 19:01:10 & 51.4 & 44.5\\
15 & 402066010 & $-1.93$ & $0.45$ & 2008-02-22 11:52:49 & 36.5 & 31.3\\
16 & 403001010 & $1.36$ & $1.05$ & 2009-02-22 19:04:19 & 71.5 & 59.7\\
17 & 403009010 & $0.17$ & $0.03$ & 2009-03-21 02:03:28 & 110.8 & 91.7\\
18 & 500005010 & $0.43$ & $-0.11$ & 2006-03-27 23:00:22 & 88.4 & 64.6 \\
19 & 500018010 & $-0.57$ & $-0.09$ & 2006-02-20 12:45:25 & 106.9 & 46.6 \\
20 & 500019010 & $-1.09$ & $-0.04$ & 2006-02-23 10:51:11 & 13.3 & 12.2 \\
21 & 501008010 & $-0.16$ & $-0.19$ & 2006-09-26 14:18:16 & 129.6 & 111.3  \\
22 & 501009010 & $-0.07$ & $0.18$ & 2006-09-29 21:26:07 & 51.2 & 47.7  \\
23 & 501010010 & $-1.29$ & $-0.64$ & 2006-10-07 02:16:52 & 50.7 & 45.7\\
24 & 501039010 & $0.78$ & $-0.16$ & 2007-03-03 12:20:20 & 96.4 & 91.1  \\
25 & 501040010 & $0.61$ & $0.07$ & 2006-09-21 17:29:01 & 61.4 & 53.9  \\
26 & 501040020 & $0.61$ & $0.07$ & 2006-09-24 05:03:12 & 44.8 & 40.0  \\
27 & 501046010 & $-0.17$ & $0.34$ & 2007-03-10 15:03:10 & 25.2 & 25.0  \\
28 & 501047010 & $-0.50$ & $0.34$ & 2007-03-11 03:55:59 & 25.6 & 19.1  \\
29 & 501048010 & $-0.83$ & $0.34$ & 2007-03-11 19:04:59 & 27.5 & 24.1  \\
30 & 501049010 & $-1.17$ & $0.33$ & 2006-10-08 10:22:40 & 19.6 & 17.6  \\
31 & 501050010 & $-0.83$ & $-0.00$ & 2006-10-09 02:20:25 & 22.0 & 18.6  \\
32 & 501051010 & $-1.17$ & $-0.00$ & 2006-10-09 13:40:09 & 21.9 & 21.1  \\
33 & 501052010 & $-1.50$ & $-0.00$ & 2006-10-10 06:45:09 & 19.3 & 16.0  \\
34 & 501053010 & $-1.83$ & $-0.00$ & 2006-10-10 21:18:59 & 21.9 & 19.9  \\
35 & 501054010 & $-0.17$ & $-0.33$ & 2007-03-12 08:11:07 & 26.1 & 23.5  \\
36 & 501055010 & $-0.50$ & $-0.33$ & 2007-03-12 23:59:09 & 27.2 & 21.2  \\
37 & 501056010 & $-0.83$ & $-0.33$ & 2007-03-13 15:41:12 & 26.5 & 25.3  \\
38 & 501057010 & $-1.17$ & $-0.34$ & 2006-10-11 10:07:27 & 20.5 & 19.1  \\
39 & 501058010 & $1.30$ & $0.20$ & 2007-03-14 05:02:29 & 63.3 & 51.1  \\
40 & 501059010 & $1.17$ & $0.00$ & 2007-03-15 18:55:51 & 62.2 & 54.4  \\
41 & 501060010 & $1.50$ & $0.00$ & 2007-03-17 05:07:04 & 64.8 & 54.6  \\
42 & 502002010 & $0.17$ & $-0.67$ & 2007-10-09 16:40:54 & 23.2 & 20.9\\
43 & 502003010 & $-0.17$ & $-0.67$ & 2007-10-10 03:41:13 & 21.5 & 18.9\\
44 & 502004010 & $0.17$ & $-1.00$ & 2007-10-10 15:21:17 & 19.9 & 18.8\\
45 & 502005010 & $-0.17$ & $-1.00$ & 2007-10-11 01:01:17 & 20.6 & 18.2\\
46 & 502006010 & $0.17$ & $0.33$ & 2007-10-11 11:34:01 & 22.6 & 21.7\\
47 & 502007010 & $0.17$ & $0.66$ & 2007-10-11 23:09:15 & 22.0 & 19.5\\
48 & 502008010 & $-0.17$ & $0.66$ & 2007-10-12 09:52:59 & 23.8 & 22.9\\
49 & 502009010 & $1.83$ & $-0.00$ & 2007-10-12 21:52:24 & 20.9 & 19.6\\
50 & 502010010 & $0.50$ & $0.33$ & 2007-10-13 07:32:00 & 21.6 & 21.2
\enddata
\end{deluxetable}

\addtocounter{table}{-1}
\begin{deluxetable}{cccccrr}
\tablewidth{0pt}
\tablecaption{
Continued.
\label{table:observation_info_gc_all_2}
}
\renewcommand{\baselinestretch}{0.85}\selectfont

\tabletypesize{\scriptsize}

\tablehead{
 &
 \colhead{Obs. ID\tablenotemark{a}} &
 \multicolumn{2}{c}{Coordinate\tablenotemark{b}} &
 \colhead{Start time} &
 \multicolumn{2}{c}{Exposure\tablenotemark{c}}\\
 & 
 \colhead{} &
 \colhead{$l$} &
 \colhead{$b$} &
 \colhead{UT} &
 \colhead{XIS} &
 \colhead{PIN}
}
\startdata
51 & 502011010 & $0.83$ & $0.33$ & 2007-10-13 18:51:09 & 23.0 & 22.1\\
52 & 502016010 & $-1.08$ & $-0.48$ & 2008-03-02 18:08:00 & 70.5 & 61.8\\
53 & 502017010 & $-0.95$ & $-0.65$ & 2008-03-06 13:26:36 & 72.6 & 64.0\\
54 & 502018010 & $-1.27$ & $-0.42$ & 2008-03-08 16:02:17 & 79.0 & 70.2\\
55 & 502020010 & $1.05$ & $-0.17$ & 2007-09-06 00:26:47 & 139.1 & 124.5  \\
56 & 502022010 & $0.23$ & $-0.27$ & 2007-08-31 12:33:33 & 134.8 & 116.8\\
57 & 502051010 & $0.92$ & $0.01$ & 2008-03-11 06:19:45 & 138.8 & 122.2  \\
58 & 502059010 & $-0.00$ & $-2.00$ & 2007-09-29 01:40:51 & 136.8 & 110.5\\
59 & 503007010 & $0.33$ & $0.17$ & 2008-09-02 10:15:27 & 52.2 & 44.2  \\
60 & 503008010 & $0.00$ & $-0.38$ & 2008-09-03 22:53:29 & 53.7 & 42.8  \\
61 & 503009010 & $-0.32$ & $-0.24$ & 2008-09-05 06:57:08 & 52.4 & 40.3  \\
62 & 503010010 & $-0.69$ & $-0.05$ & 2008-09-06 15:56:13 & 53.1 & 37.1  \\
63 & 503011010 & $-0.97$ & $-0.13$ & 2008-09-08 09:08:09 & 57.6 & 40.2  \\
64 & 503012010 & $-0.91$ & $-0.45$ & 2008-09-14 19:35:07 & 57.7 & 51.9  \\
65 & 503013010 & $-1.30$ & $-0.05$ & 2008-09-16 00:51:19 & 104.8 & 93.9  \\
66 & 503014010 & $-2.10$ & $-0.05$ & 2008-09-18 04:46:49 & 55.4 & 51.2  \\
67 & 503015010 & $-2.35$ & $-0.05$ & 2008-09-19 07:33:05 & 56.8 & 52.8  \\
68 & 503016010 & $-2.60$ & $-0.05$ & 2008-09-22 06:47:49 & 52.2 & 49.3  \\
69 & 503017010 & $-2.85$ & $-0.05$ & 2008-09-23 08:08:10 & 51.3 & 48.6  \\
70 & 503021010 & $-1.62$ & $0.20$ & 2008-10-04 03:44:03 & 53.8 & 49.6  \\
71 & 503072010 & $-0.42$ & $0.17$ & 2009-03-06 02:39:12 & 140.6 & 135.5\\
72 & 503076010 & $-1.50$ & $0.15$ & 2009-02-24 17:04:51 & 52.9 & 43.8\\
73 & 503077010 & $-1.70$ & $0.14$ & 2009-02-26 01:01:00 & 51.3 & 43.7\\
74 & 503081010 & $0.03$ & $-1.66$ & 2009-03-09 15:41:50 & 59.2 & 57.6\\
75 & 503099010 & $-0.22$ & $1.13$ & 2009-03-10 19:39:08 & 29.7 & 30.6 \\
76 & 503100010 & $-0.69$ & $1.13$ & 2009-03-15 06:41:41 & 25.7 & 24.1 \\
77 & 503101010 & $-0.45$ & $0.89$ & 2009-03-16 14:43:17 & 33.9 & 30.8 \\
78 & 503102010 & $-0.70$ & $0.66$ & 2009-03-17 07:49:09 & 33.7 & 30.1 \\
79 & 503103010 & $-0.01$ & $1.20$ & 2009-03-11 10:56:59 & 18.3 & 16.4 \\
80 & 504050010 & $0.10$ & $-1.42$ & 2010-03-06 03:55:37 & 100.4 & 80.5\\
81 & 504088010 & $-0.00$ & $-0.83$ & 2009-10-14 11:30:56 & 47.2 & 32.6  \\
82 & 504089010 & $-0.05$ & $-1.20$ & 2009-10-09 04:05:59 & 55.3 & 40.2  \\
83 & 504090010 & $-1.49$ & $-1.18$ & 2009-10-13 04:17:20 & 41.3 & 35.0  \\
84 & 504091010 & $-1.50$ & $-1.60$ & 2009-09-14 19:37:36 & 51.3 & 47.8  \\
85 & 504092010 & $-1.44$ & $-2.15$ & 2009-09-16 07:21:35 & 50.9 & 45.6  \\
86 & 504093010 & $-1.50$ & $-2.80$ & 2009-09-17 13:54:31 & 53.2 & 46.9  \\
87 & 903004010 & $-2.75$ & $-1.84$ & 2008-10-07 16:19:21 & 15.7 & 28.7 \\
88 & 904002010 & $1.02$ & $2.53$ & 2009-08-28 12:20:31 & 23.1 & 21.9 \\
89 & 904002020 & $1.02$ & $2.53$ & 2009-09-06 19:38:32 & 25.1 & 18.8 \\
90 & 504003010 & $-1.45$ & $-0.87$ & 2010-02-25 04:33:17 & 50.9 & 41.3  \\
91 & 504001010 & $-1.47$ & $-0.26$ & 2010-02-26 09:15:00 & 51.2 & 42.2  \\
92 & 504002010 & $-1.53$ & $-0.58$ & 2010-02-27 16:14:41 & 53.1 & 46.6 
\enddata
\tablenotetext{a}{Observation ID.}
\tablenotetext{b}{Aim point in Galactic coordinate (degree).}
\tablenotetext{c}{Net exposure in units of $10^{3}$~s.}
\end{deluxetable}


\end{document}